\documentclass[10pt,aps,prb,twocolumn,superscriptaddress,floatfix,showpacs]{revtex4-1}
\usepackage{graphicx} 
\usepackage{epstopdf} 
\usepackage{amsmath}
\usepackage{color} 
\usepackage{comment}

\begin{document}

\title{Thermoelectric properties of atomic-thin silicene and germanene nano-structures}

\author{K. Yang} 
\affiliation{Nano-bio spectroscopy group and ETSF Scientific
Development Center, Departamento de Fisica de Materiales, Universidad del Pais
Vasco UPV/EHU, Avenida Tolosa 72, E-20018 San Sebastian, Spain}

\author{S. Cahangirov} 
\affiliation{Nano-bio spectroscopy group and ETSF Scientific
Development Center, Departamento de Fisica de Materiales, Universidad del Pais
Vasco UPV/EHU, Avenida Tolosa 72, E-20018 San Sebastian, Spain}

\author{A. Cantarero} 
\affiliation{Instituto de Ciencia de Materiales, Universidad de Valencia, E-46071 Valencia, Spain}

\author{A. Rubio}
\affiliation{Nano-bio spectroscopy group and ETSF Scientific
Development Center, Departamento de Fisica de Materiales, Universidad del Pais
Vasco UPV/EHU, Avenida Tolosa 72, E-20018 San Sebastian, Spain}

\author{R. D'Agosta} 
\email{roberto.dagosta@ehu.es}
\affiliation{Nano-bio spectroscopy group and ETSF Scientific
Development Center, Departamento de Fisica de Materiales, Universidad del Pais
Vasco UPV/EHU, Avenida Tolosa 72, E-20018 San Sebastian, Spain}
\affiliation{IKERBASQUE, Basque Foundation for Science, E-48011, Bilbao, Spain} 

\date{\today}

\begin{abstract} 
The thermoelectric properties in one- and two-dimensional silicon and germanium
structures have been investigated using first-principle density functional
techniques and linear response for the thermal and electrical transport. We
have considered here the two-dimensional silicene and germanene, together with
nano-ribbons of different widths. For the nano-ribbons, we have also
investigated the possibility of nano-structuring these systems by mixing
silicon and germanium. We found that the figure of merit at room temperature of
these systems is remarkably high, up to 2.5.
\end{abstract} 
\maketitle

\section{Introduction}
Thermoelectric energy conversion is the ability of a device to convert a steady
temperature gradient into an electrical current, and it was firstly discovered
by Seebeck in 1821.\cite{Pollock1985,Nolas2001,Goldsmid2010} In a reverse mode
operation, a thermoelectric device can be used as a cooler by maintaining a
steady current in the device (Peltier
effect).\cite{Pollock1985,Nolas2001,Goldsmid2010} Recently, the quest for a
highly efficient thermoelectric device has attracted tremendous interests due
to significant potential industrial
applications.\cite{Pollock1985,DiSalvo1999,Nolas2001,Vining2009,Goldsmid2010}
The efficiency of the thermoelectric conversion is characterized by a
dimensionless parameter, called figure of merit
\begin{equation}
	ZT=\frac{\sigma S^2 T}{\kappa},
	\label{figureofmerit}
\end{equation}
where $\sigma$ is the electric conductance, $S$ is the Seebeck coefficient, $T$
is the absolute temperature, $\kappa=\kappa_e+\kappa_p$ is the total thermal
conductance that is usually splits into the electron and phonon contributions,
respectively.\cite{Pollock1985,Nolas2001,Goldsmid2010} Generally speaking,
materials with $ZT\approx 1$ are regarded as good thermoelectric components,
while devices with a $ZT$ approaching to or larger than 3 could efficiently
compete with conventional energy conversion techniques. State of the art values
for the figure of merit are about 1, while higher values have been reported in
the literature for particular materials which, however, have presently proven
difficult to integrate into our technologies or to produce
industrially in a reliable way, or whose cost makes them unaffordable at large
scale.\cite{Venkatasubramanian2001} Admittedly, the optimization of the figure
of merit is a difficult problem. Indeed, an ideal thermoelectric material
should hold the electric conductance and the Seebeck coefficient as high as
possible, while keeping the thermal conductance as low as possible.
Unfortunately, because of the Wiedemann-Franz law $\kappa_e/\sigma=(k_B \pi)^2
T/3e^2$ (valid in a great extent for metals), where $k_B$ and $e$ are,
respectively, the Boltzmann constant and carrier charge,\cite{Ashcroft1976} the
two conductances are locked together and increasing the first leads to an
increase in the second. It therefore looked natural to attempt to decrease the
phonon thermal conductance since this will hopefully not (strongly) affect the
electronic properties, although the maximum $ZT$ achieved so far makes these
devices not commercially viable.

After the seminal work by Hicks and Dresselhaus,\cite{Hicks1993} a strong
research activity has been focused on nano-structured materials for
thermoelectric applications. This boost can easily be explained as an attempt
to escape from the Wiedemann-Franz law while dramatically increasing the
electronic density of states.\cite{Minnich2009,Pichanusakorn2010} With the
discovery of graphene,\cite{Novoselov2004} and the subsequent investigation of
its properties, it became apparent that graphene is not an efficient
thermoelectric material since its thermal conductance is extremely high.
\cite{Balandin2008,Seol2010,Prasher2010} On the other hand, it has been shown
that nano-structuring graphene with boron-nitride in a nano-ribbon increased
the overall figure of merit by a factor 20.\cite{Yang2012} Notwithstanding its
phenomenal properties, the integration of graphene with the actual
silicon-based technologies has proven a quite challenging task, whose solution
would probably require the complete redesign of electronics devices. As our
present technology is based on silicon (Si) and germanium (Ge) semiconductors,
it thus appears natural to look at the thermoelectric properties of these
materials, since the integration of a thermoelectric device based on them would
be simpler than the integration of carbon-based devices. For example, in
silicon nano-wires, the thermal conductance can be reduced in a factor of 100
due to the quenching of phonon transport and they exhibit a high thermoelectric
conversion ratio.\cite{Hochbaum2008} This suggests a prospective avenue to
improve the thermoelectric performance through decreasing the characteristic
size of materials and various nano-structures such as nano-tubes and
nano-membranes can be proposed.

Silicene resembles graphene
\cite{CastroNeto2009,Peres2010,Ayala2010,DasSarma2011,Kotov2012} in the atomic
single layer arrangements, i.e., it forms a honeycomb lattice and shares with
the carbon system similar electronic properties. In particular, it is viewed as
new type of atomic-layered materials with outstanding properties such as the
zero effective mass at the Dirac-point and infrared absorbance optical
spectra.\cite{Cahangirov2009,Chen2012,Bechstedt2012} Experimentally, single
layer silicene (buckled)\cite{DePadova2010,Lalmi2010,DePadova2012,Kara2012,
Feng2012,Enriquez2012,Jamgotchian2012,
Vogt2012,Lin2013,Chen2013,Meng2013,Cahangirov2013} and silicene nano-ribbons
(SiNRs)\cite{DePadova2010,DePadova2012} have been synthesized on Ag substrate.
In particular, SiNRs up to a narrow width of 1.6 nm have been produced, aligned
parallel to each other in a well-distributed way.\cite{DePadova2010} From the
experience gained with the current micro-electronics, we know that Ge is a good
partner for Si since they share similar electronic properties and form bulk
crystal with comparable lattice constant ($a_{\rm Si}=0.5431$ nm, while $a_{\rm
Ge}=0.5658$ nm, with a lattice mismatch of 4\%). The elastic limit is around
7\%. In the case of InAs, for instance, only one monolayer can be grown on
GaAs.\cite{Li1994} A single layer hexagonal lattice of Ge, called germanene,
has been predicted from an ab-initio calculations.\cite{Cahangirov2009}
Theoretically, germanene presents a Dirac point, and the electronic and
structural properties of this material would be very similar to those of
silicene. We will discuss some of them in more detail in the following. In
particular, we will consider germanene nano-ribbons (GeNRs) of different widths
and the possibility of forming Si-Ge single atomic layered nano-ribbons by
alternating stripes of Si with stripes of Ge.

In this paper, we investigate with ab-initio technique combined with linear
response approach the thermoelectric properties of both two-dimensional (2D)
silicene and germanene nano-sheets and one-dimensional (1D) nano-ribbons. We
find that some of these systems will show a figure of merit larger than 1 at
room temperature (with a maximum value of 2.18). Our results are consistent
with those obtained by Pan et al.\cite{Pan2012}, although they are based on
different techniques especially for the calculations of the phonon thermal
conductance. We believe that this agreement is partially fortuitous, as we will
discuss in the following.
  
The paper is organized as follows. In Sec. \ref{Method} we will discuss in
details the numerical and theoretical methods we used to investigate the Si and
Ge systems. In Sec. \ref{2D}, we introduce the 2D systems, silicene and
germanene, study their stability and investigate their transport properties. In
Sec. \ref{1D}, we investigate the 1D nano-ribbons. In this section we focus
mostly on the Ge system, since the SiNRs have been investigated elsewhere, and
from our calculations, Si and Ge nano-ribbons do share essentially some same
properties. We find that the nano-ribbons can have a quite large figure of
merit. This is due to the fact that both Si and Ge nano-ribbons do have a
finite electronic gap that dramatically enhances the Seebeck coefficient. In
Sec.~\ref{1DSiGe}, we consider nano-ribbons created by alternating stripes of
Si and Ge. By nano-structuring the nano-ribbons we would like to confine the
phonons and therefore decrease the thermal conductance. However, we report that
the SiGeNRs do show some the same transport properties of the pure Si or pure
Ge nano-ribbons. This is due to the limitations of our method of choice, namely
a full ab-initio study for the phonon energy transport. Indeed, within this
technique we are limited to fairly small nano-ribbons and therefore the
long-wavelength phonons are not quenched by the regular pattern of the
structured nano-ribbons. On the other hand, a classical technique, based on
molecular dynamics, would allow us to calculate the thermal conductance of
larger devices. However, this technique does not recover the correct quantum
limit of these one-dimensional systems, and therefore we do expect that the
molecular dynamics results to give the incorrect thermal conductance at
temperature below the Debye temperature, which for Si and Ge systems can be
estimated to be about 640 K and 374 K, respectively. We check the idea that
nano-structuring would decrease the phonon thermal conductance by using a
tight-binding approximation, which allows us to consider larger super-cell than
a purely ab-initio method. We indeed show that the thermal conductance does
strongly decrease when we consider a hetero-structure of Si and Ge. Finally, in
Sec. \ref{Conclusions} we draw our conclusions and some outlooks of this work.

\section{Method}
\label{Method}

In linear response theory, by using the Onsager's relations and the
Landauer's theory of quantum transport, the electrical conductance $\sigma$, the
Seebeck coefficient $S$, and the electron contributed thermal conductance
$\kappa_e$, can be written as\cite{Nolas2001,Goldsmid2010,DAgosta2013}
\begin{eqnarray} 
	&&\sigma_{\alpha \beta}(\mu,T)=e^2 L_{00}(\mu,T),\label{conductance}\\
	&&S_{\alpha \beta}(\mu,T)=\frac{1}{eT}\frac{L_{01}(\mu,T)}{L_{00}(\mu,T)},\label{seebeck}\\
	&&\kappa_{e,\alpha \beta}(\mu,T)=\frac{1}{T} \left[L_{11}(\mu,T) +
	\frac{L_{01}(\mu,T)^2}{L_{00}(\mu,T)}\right]\label{thermalc}, 
\end{eqnarray} 
where 
\begin{equation}
	L_{mn}(\mu,T)= -\frac{1}{A} \int^{\infty}_{-\infty} {d\epsilon\, {\cal
	T}_{e,\alpha \beta}(\epsilon) (\epsilon-\mu)^{m+n} \frac{\partial
	f(\epsilon,\mu,T)}{\partial \epsilon}}
\end{equation} 
is the Lorenz integral. In these equations, $\mu$ is the chemical potential,
$A$ is the area of the considered system, $\alpha$ and $\beta$ are the indices of the
spatial components, $x$, $y$ and $z$, $f(\epsilon,\mu,T)$ is the Fermi
distribution function at a given temperature $T$, and $\mathcal{T}_e$ is a
transmission function which is related to the probability of electrons to
cross the system.\cite{Landauer1957,Baranger1989} Similarly, the phonon thermal
conductance is given by\cite{Yamamoto2006}
\begin{equation} 
	\kappa_{p,\alpha \beta}(T)=\frac{1}{A} \int^{\infty}_{0} {d\omega\, 
	{\cal T}_{p,\alpha
	\beta}(\omega) \hbar \omega \frac{\partial n(\omega,T)}{\partial T}},
\end{equation} 
where $\omega$ is the phonon-vibrational frequency, $\hbar$ is the reduced
Planck-constant, and $n(\omega,T)$ is the Bose-Einstein distribution function.
Again, ${\cal T}_p$ is a transmission function for phonons. A common
expression of the electron and phonon transmission functions can be given in
terms of the electron and phonon band structures, respectively,
\begin{equation}
	{\cal T}_{e/p, \alpha \beta}(E)=\frac{1}{N} \sum_{i,\mathbf{k}}
	\tau_{e/p,i,\mathbf{k}} \upsilon_\alpha(i,\mathbf{k})
	\upsilon_\beta(i,\mathbf{k}) D_{e/p}(E_{i,\mathbf{k}}),
	\label{transmission}
\end{equation} 
where $N$ is the number of sampled $\mathbf{k}$-point in the first Brillouin-zone,
$i$ is the band index, $\tau_{e/p}$ is the relaxation time of
electrons/phonons, $\upsilon$ is the velocity calculated from the band
dispersion, and $D_{e/p}(E_{i,\mathbf{k}})$ is the electron/phonon density of states
associated with the band $i$.

To obtain the energy band structure, we perform first-principle calculations
within the local density approximation by using the projector-augmented wave
potentials as implemented in VASP.\cite{Kresse1996} The exchange correlation
energy is chosen in the form of Ceperley-Alder which has been parameterized by
Perdew and Zunger.\cite{Ceperley1980,Perdew1981} For the self-consistent
potential and the total energy calculations, the $\mathbf{k}$-points of the
Brillouin-zone in the reciprocal space are sampled by a (25$\times$1$\times$1)
Monkhorst-Pack grids. The kinetic energy cut-off is set to 500 eV. After ionic
relaxation, the Hellmann-Feynman forces acting on each atom are less than 0.01
eV/{\AA}. We obtain the force constant matrix for the calculation of the phonon
dispersion, through the small displacement method.\cite{Alfe2009} We use a
supercell technique with a 15 {\AA} of vacuum. In these calculations we have
neglected both the phonon-phonon and the electron-phonon interactions. We
expect that for the low energy phonons, mostly responsible for the thermal
transport, the correction due to these interactions to be small, especially for
the Seebeck coefficient.

In the following, we will consider both the figure of merit of Eq. (\ref{figureofmerit})
and the electronic figure of merit $ZT_e$, defined as
\begin{equation}
	ZT_e=\frac{S^2\sigma}{\kappa_e}T.
	\label{zte}
\end{equation}
Then Eq.~(\ref{figureofmerit}) can be rewritten as
\begin{equation}
	ZT=\frac{S^2\sigma}{\kappa_p+\kappa_e}T
	=\frac{S^2\sigma}{\kappa_e}T\left(\frac{1}{1+\kappa_p/\kappa_e}\right)
	=\frac{ZT_e}{1+\kappa_p/\kappa_e}.
	\label{zt}
\end{equation}

Although $ZT_e$ is not a physical measurable quantity, it is useful because it
provides an upper bound to the total figure of merit, and since it does not
include the phonon thermal contribution is easier to calculate. A small $ZT_e$
will therefore imply a small figure of merit $ZT$.

\section{Two-dimensional silicon and germanium crystals}
\label{2D}

\begin{figure}[ht!] 
	\includegraphics[width=8.6cm]{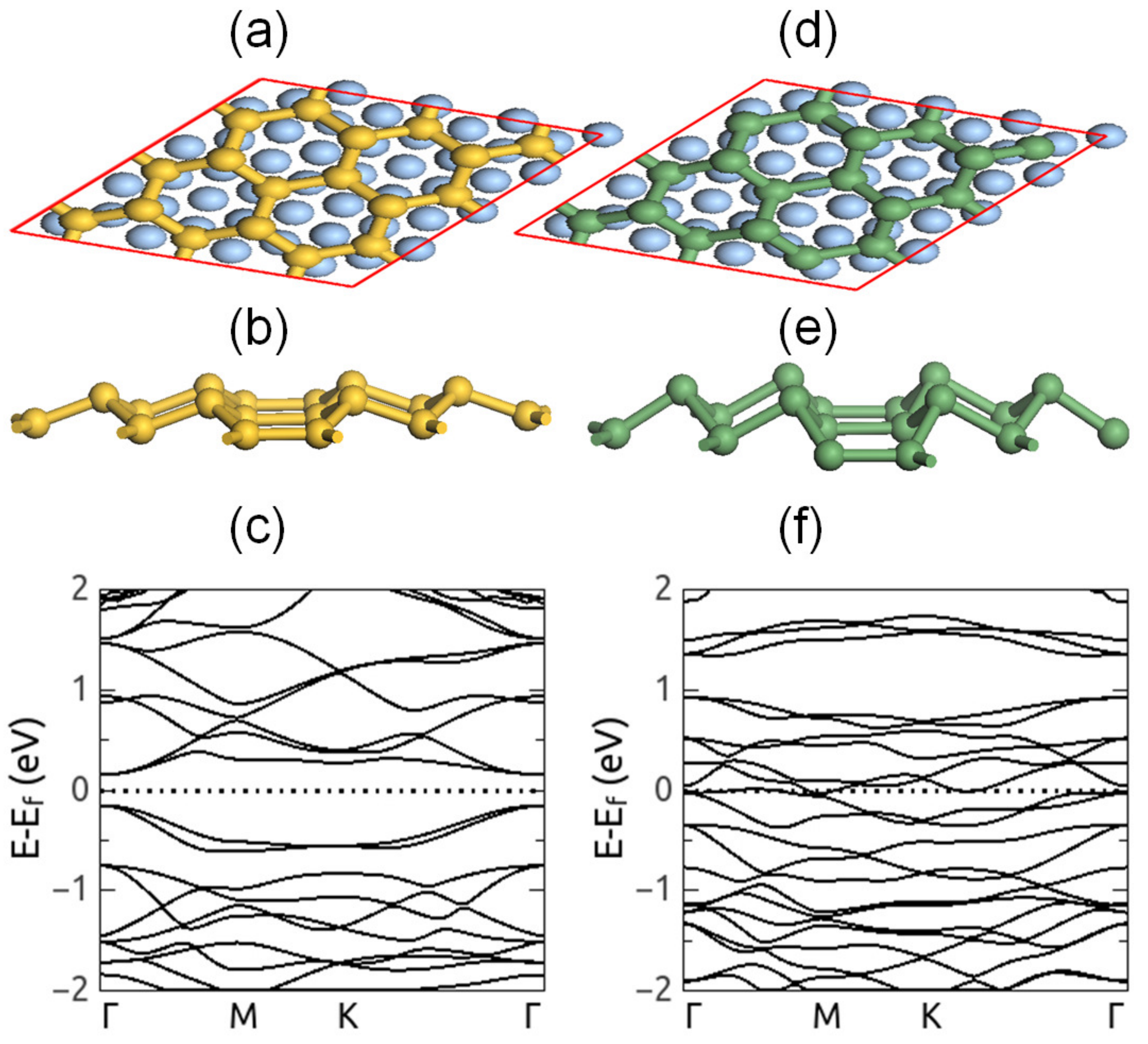} 
	\caption{(Color online) Geometrical structures of one-layer 3$\times $3 (a)
silicene and (d) germanene on top of five-layers 4$\times$4 Ag(111). (b) and
(e) Distorted silicene and germanene obtained by removing the silver substrate
from (a) and (d), respectively. (c) and (f) Electronic energy bands
corresponding to the distorted silicene and germanene grown on Ag(111), 
respectively, where the dotted line denotes the Fermi energy.}
	\label{fig1} 
\end{figure}

We first investigate the electronic properties of a single-layer of silicon,
i.e., silicene. In trying to closely reproduce the experimental
setup,\cite{Vogt2012,Feng2012,Lin2013} we put one-layer of 3$\times$3 silicene
on top of five-layers of 4$\times$4 Ag(111): according to experimental
evidence, the two lattices should match, thus decreasing the total stress at
the boundary and creating an ideal supercell for our calculations. The
geometrical structure for silicene obtained after the full relaxation is shown
in Fig. \ref{fig1}(a) and corresponds to the structure discussed in
Ref.~\onlinecite{Cahangirov2013}. We have superimposed the Ag(111) layer to
show the excellent structural matching, as highlighted by the boundary
continuous (red) line. Figure 1(b) shows the silicene obtained by removing the
silver substrate in Fig. \ref{fig1}(a). Contrary to graphene, silicene is not a
strict two-dimensional system, in the sense that the atoms in silicene are
arranged on two atomic layers with a fairly small buckling distance, which
depends on the presence of the substrate. Indeed, it is found that the atomic
arrangement is further distorted by the metallic
substrate.\cite{Cahangirov2013} Starting from a single layer of silicon,
arranged in a plane on an hexagonal lattice without the Ag substrate, we would
have obtained a system with a different buckling, where the atoms would divide
equally between the upper and lower plane. In our optimized structure however,
we observe that the silicene presents buckling forming two atomic layers with
six atoms on top of the other twelve atoms which are therefore closer to the Ag surface.
The buckling distance between these two layers is about 0.79 {\AA}. In Fig.
\ref{fig1}(c) the electronic energy band for the distorted silicene is plotted
along the high-symmetry points of the first Brillouin-zone, where the dotted
line indicates the Fermi energy that we set for convenience at 0. It can be
seen that a band gap about 0.3 eV crosses the Fermi energy, indicating
semiconducting properties of the system. This must be compared with the flat
silicene (unoptimized structure) and the silicene optimized without the Ag
substrate, which both present a Dirac point at the K point of the first
Brillouin-zone, therefore both showing metallic properties (see Fig.
\ref{fig3}). A detailed discussion of the electronic structure of supported
silicene can be found in Ref.~\onlinecite{Cahangirov2013}.

Germanene is an analog of silicene, where the silicon atoms are replaced by
germanium. Although, to the best of our knowledge, up to now there is no direct
experimental observation of these structures, here we study the electronic
properties of two-dimensional germanene. Figure 1(d) shows the atomic structure
of one-layer 3$\times$3 germanene on top of five-layers 4$\times$4 Ag(111), and
Fig.~\ref{fig1}(e) shows the unsupported single-layer germanene by removing the
silver substrate. The structure has been fully relaxed. It is found from Fig.
\ref{fig1}(e) that similar to silicene, two layers are formed with six Ge atoms
on the top layer and the other twelve Ge atoms on the bottom layer closer to
the Ag surface. The buckling distance between the two layers is about 1.42
{\AA}. Figure \ref{fig1}(f) shows the band structure of the distorted germanene
without the Ag substrate. It is found that there is no gap through the Fermi
energy, indicating metallic properties. The zero gap observed in germanene
originates from the high-buckling distance between the two atomic layers.

\begin{figure}[ht!]
	\includegraphics[width=8.6cm]{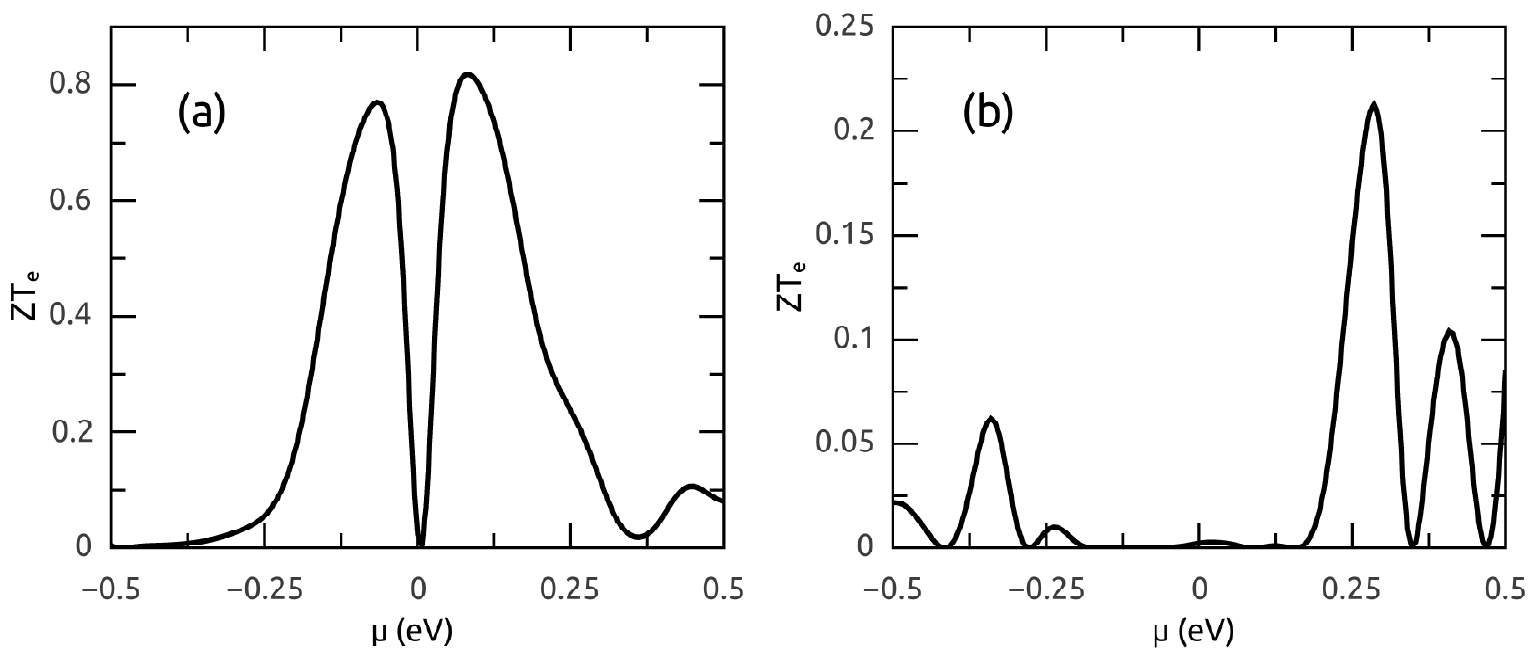} 
	\caption{(a) and (b) Dimensionless electronic figure of merit $ZT_e$ at room temperature
as a function of chemical potential $\mu$ corresponding to the unsupported distorted silicene
and germanene, respectively.}
	\label{fig2}
\end{figure}

\begin{figure}[ht!] 
	\includegraphics[width=8.6cm]{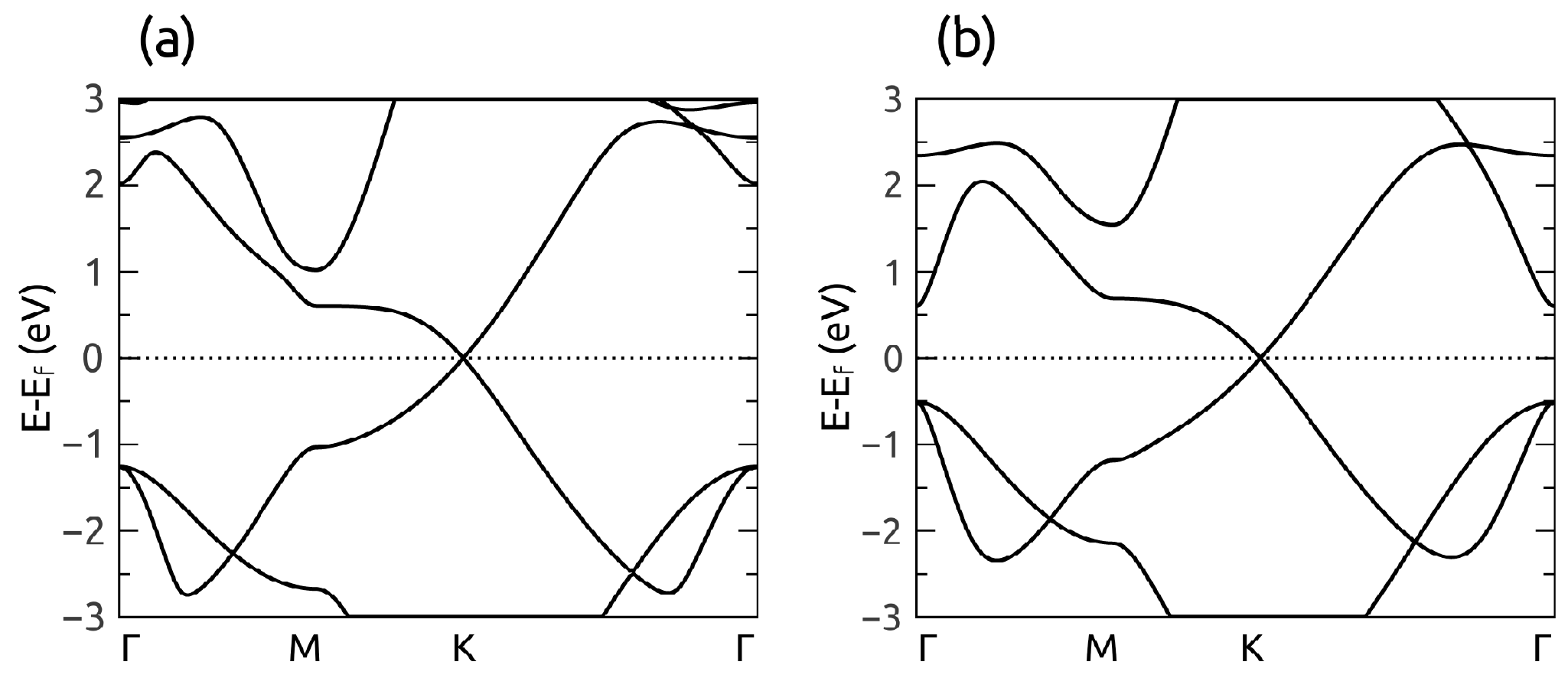}
	\caption{Electron energy bands of free standing (a) silicene
and (b) germanene, respectively, where the dotted line denotes the Fermi energy.}
	\label{fig3} 
\end{figure}

Based on the energy bands, we calculate the thermoelectric coefficients of the
two-dimensional silicene and germanene structures at room temperature,
$T=300~\mathrm{K}$. To perform the calculation of the figure of merit we have
evaluated the transport coefficients given in
Eqs.~(\ref{conductance})-(\ref{thermalc}) in the constant relaxation time
approximation. We have used the BoltzTraP code\cite{Madsen2006} to perform the
integration in the first Brillouin-zone obtained from the VASP calculations. As
we discussed in the introduction, we provide an upper bound to $ZT$, in the
form of the electronic figure of merit $ZT_e$. We notice that the electronic
figure of merit $ZT_e$ is independent of the relaxation time, thus justifying
the use of the BoltzTraP. Figures \ref{fig2}(a) and (b) show the dimensionless
electronic figure of merit, $ZT_e$ as a function of the chemical potential
$\mu$ for the distorted silicene and germanene, respectively. It can be seen
that the figure of merit for silicene exhibits two peaks in the left- and
right-hand sides of $\mu=0$, which separately correspond to the hole and
electron transport. The maximum of the peak is about $ZT_e=0.81$, while for the
unsupported germanene, it can be seen from Fig. \ref{fig2}(b) that the peak of
$ZT_e$ is very small at $\mu=0$, although some peaks appear at $\sim\pm 0.3$
eV. The reason is that the unsupported germanene has a metallic character which
leads to a very small Seebeck coefficient.

In Fig. \ref{fig3} we show the electronic properties of free standing silicene
and germanene. After the full relaxation, it is found that the buckling distance
for silicene is about 0.43 {\AA},
and for germanene 0.65 {\AA}. For both the free standing silicene and
germanene, from Figs. \ref{fig3}(a) and (b) it can be seen that
there is no gap at the Fermi energy. Indeed, at
the high symmetry point K, a linear energy dispersion is shown in the band
structures, indicating the existence of the massless Dirac fermions in these low dimensional Si structures\cite{Cahangirov2009} similar to the
graphene.\cite{CastroNeto2009}  

\begin{figure}[ht!]
	\includegraphics[width=8.6cm]{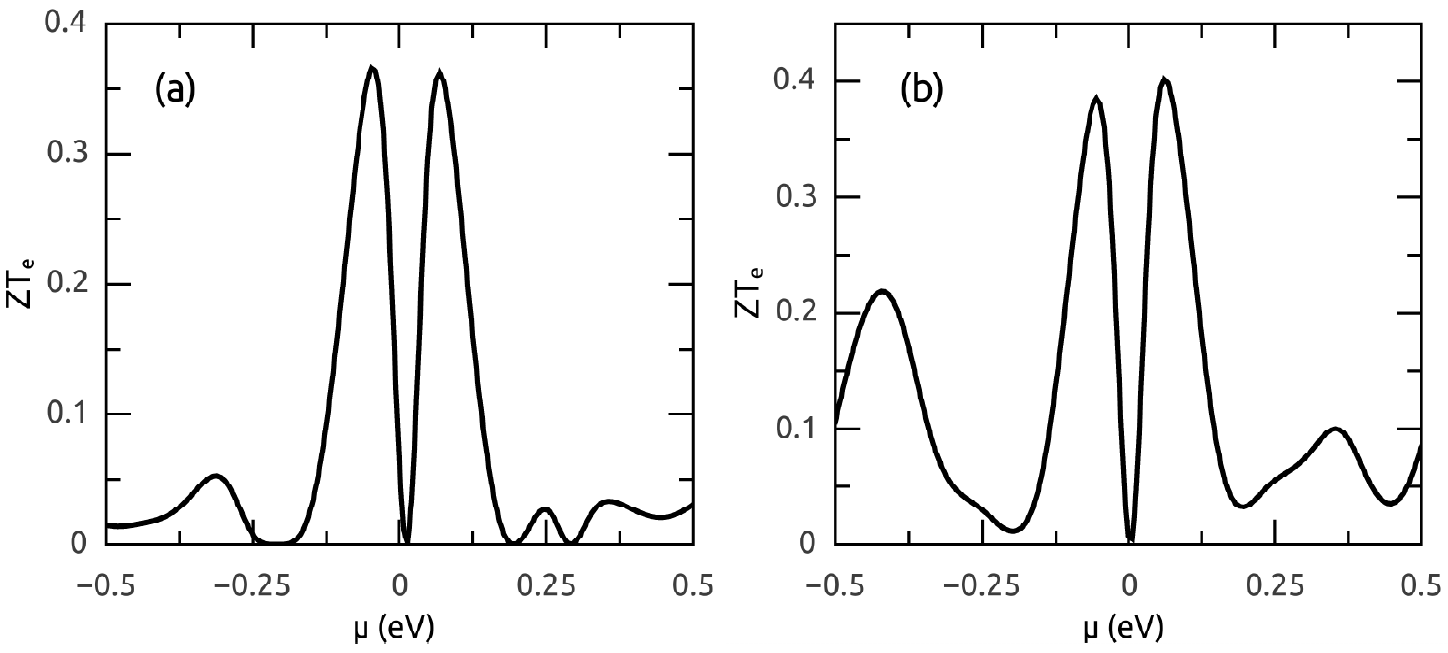}
	\caption{(a) and (b) Dimensionless electronic figure of merit $ZT_e$ at room temperature
as a function of chemical potential $\mu$ for the free standing silicene
and germanene, respectively.}
	\label{fig4}
\end{figure}

Through the energy band structure calculations, in Fig. \ref{fig4} we investigate the
dimensionless electronic figure of merit $ZT_e$ for both free standing silicene and
germanene. It is found that the figure of merit for silicene and germanene shows
two peaks near $\mu=0$. The maximum value of $ZT_e$ is 0.36 (see Fig. \ref{fig4}(a)),
while the maximum of the peak for germanene is 0.41 (see Fig. \ref{fig4}(b)). 

We have shown that silicene and germanene, crystal structures similar to
graphene where carbon is replaced by either silicon or germanium, might
possibly have a figure of merit of the order of 1. Our calculations provide an
upper limit to the theoretical figure of merit since in these calculations we
are not including the phonon thermal conductance and suggest that silicene
might have a better thermoelectric properties in this 2D system
since it does present a gap in the electronic energy spectrum which
corresponds to a large Seebeck coefficient.

\section{Quasi-one-dimensional nano-structures} 
\label{1D}

We now consider quasi-one-dimensional systems, nano-ribbons, made of stripes of
germanene or silicene of finite width. We assume that it is possible to “cut”
those stripes from the respective crystal by removing the excess material. It
has been reported that SiNRs can have a quite large figure of
merit, up to 5 at 600 K.\cite{Pan2012} Motivated by these results, and by the expectation than germanene nano-ribbons might perform better since their Debye temperature is lower, we have
investigated the thermoelectric efficiency of GeNRs and, in
the next session, nano-ribbons obtained by alternating Si and Ge
nano-structures or by randomising the Si and Ge arrangements. As
standard with nano-ribbons, there are two ways to terminate the edges of the
ribbons (see Fig.~\ref{fig5}), forming either zigzag or armchair edges. We
identify the quantities associated with the zigzag with a Z and those of the
armchair with a A. As to these one-dimensional systems,
the electrical conductance $\sigma$, Seebeck coefficient $S$,
and thermal conductance could be contracted into a scalar instead of a tensor, and
the $\alpha$ and $\beta$ are fixed in the $x$ direction (see details in Ref.~\onlinecite{Yang2012}).
Therefore in this case we can calculate the electron and phonon transmissions ${\cal T}_{e/p}$
by counting the number of transport modes from the energy band.
This gives the same results by comparing them
with the constant relaxation time approximation in Eq. (\ref{transmission})
except by an amplitude factor difference.

\subsection{Germanene nano-ribbons}

\begin{figure}[ht!]
	\includegraphics[width=8.6cm]{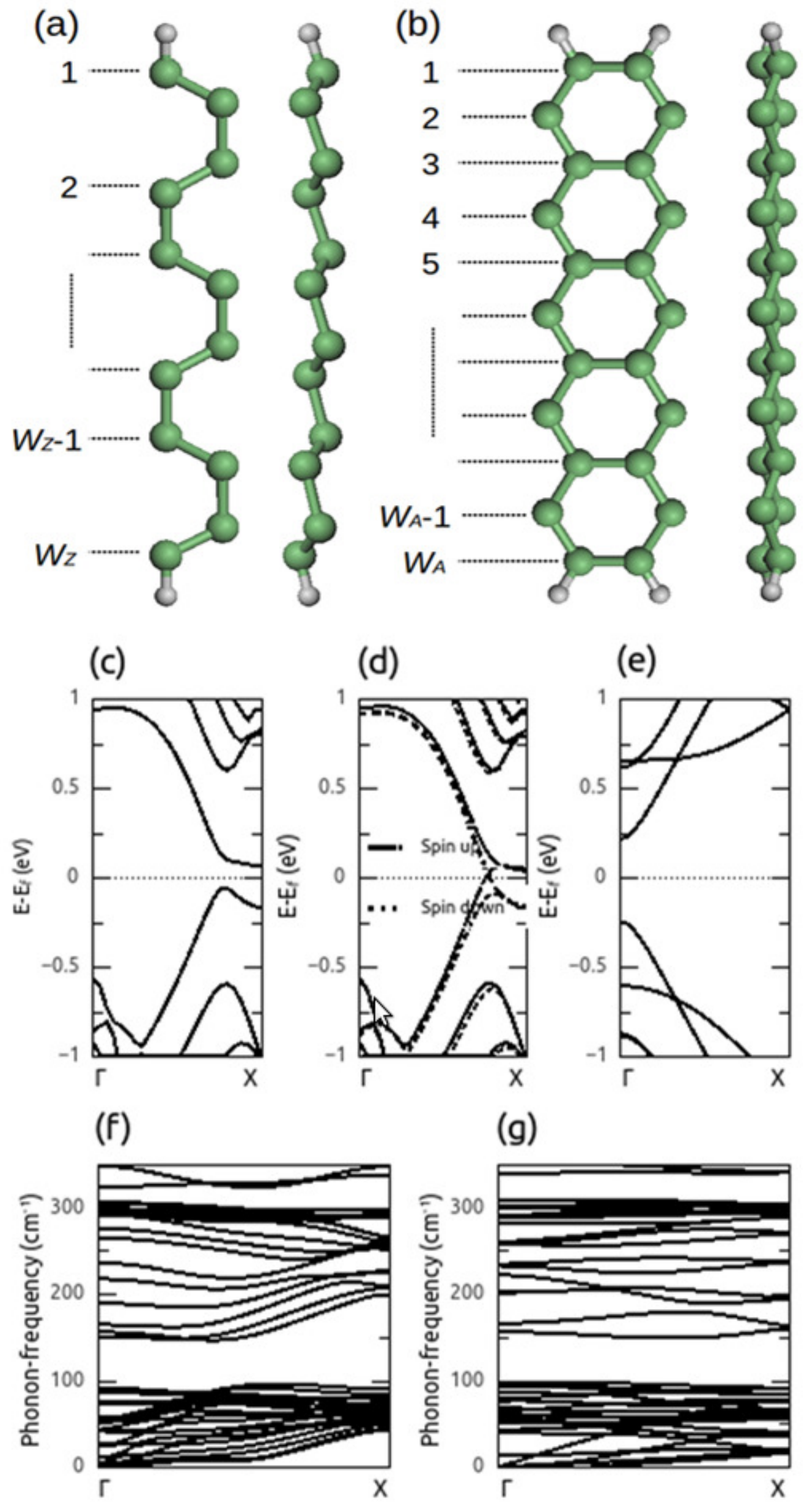}
	\caption{(Color online) (a) and (b): Optimized geometrical structures of
Z-GeNRs and A-GeNRs and their lateral views. For the atoms at the edges, we
passivate the unsaturated bonds with hydrogen atoms. $W_{Z}$ and $W_{A}$ denote
the width of the nano-ribbons for the zigzag and armchair terminated
nano-ribbons, respectively. (c) Electron energy band for Z-GeNRs with $W_{Z}=6$
for the AFM state. Notice the presence of a small electronic gap.
(d) Electron energy band for Z-GeNRs with $W_{Z}=6$ for the FM state.
(e) Electron energy band of A-GeNRs with $W_{A}=6$ corresponding
to the NM state. In (c), (d) and (e) the Fermi energy is chosen as the
reference energy and set to 0. (f) and (g): Phonon energy dispersions for Z-GeNRs with
$W_{Z}=6$ and A-GeNRs with $W_{A}=6$, respectively.}
	\label{fig5} 
\end{figure}

Figures \ref{fig5}(a) and (b) show the optimized structures of zigzag-
and armchair-edged GeNRs (Z-GeNRs and A-GeNRs),
respectively. To see more clearly the buckling, we report here a side view of these structures.
Hydrogen atoms are used to passivate the unsaturated bonds of
the Ge atoms at the edges. $W_{Z}$ and $W_{A}$ identify the ribbon
width. It can be seen from the top view that GeNRs form hexagonal rings as the
union of two sub-lattices, but, at odds with what happens for graphene
nano-ribbons, atoms in these two sub-lattices do not belong to the same plane:
in the vertical direction there is some buckling, which is almost uniform for
the atoms at the edge or in the center. Our calculations give for the Z-GeNRs a
buckling distance of 0.62 {\AA}, while for A-GeNRs give 0.66 {\AA}. For these
nano-ribbons, our total energy calculations show that the anti-ferromagnetic
(AFM) state of Z-GeNRs is more stable than the ferromagnetic (FM) and
non-magnetic (NM) states counterparts. This is in agreement with other
calculations performed for SiNRs\cite{Pan2012} and theoretical predictions
originally derived for graphene, which we expect to be valid for these
systems.\cite{Jung2009} However, the energy difference between the different
magnetic phases is small. This might be important for device stability
especially at temperatures higher than 300 K. The bands of AFM and FM states
are shown in Figs. \ref{fig5} (c) and (d), respectively, where the dotted line
corresponds to the Fermi energy. We can see that AFM state exhibits a finite
small gap: The bands for spin up and down are degenerate and the gap is about
0.1 eV. While for the FM state, it is found that spin up and down are
non-degenerate, producing metallic properties, and similar properties are valid
for the NM state (not shown). In the case of A-GeNR, our calculations indicate
that the NM state is stable, indicating semiconducting properties as shown in
Fig. \ref{fig5} (e). Because the metallic system produces bad thermoelectric properties
(generally the $ZT$ is smaller than 0.1),
in the rest part of this work, we will focus the attentions on the AFM state in the
zigzag-edged nano-ribbons and NM state in the armchair-edged nano-ribbons. To
confirm the structural stability of GeNRs, we have calculated the phonon
dispersion relations. In Figs.~\ref{fig5} (f) and (g) we report the phonon dispersion
relation for the nano-ribbons with width 6 for both Z-GeNR and A-GeNR,
i.e., $W_{Z}=W_{A}=6$, respectively. It can be seen that in the limit of
$\omega \rightarrow 0$, there are four acoustic phonon modes in the spectrum
stemming from the lattice symmetry. In particular, no negative phonon mode is
observed, which confirms that both the Z-GeNRs and A-GeNRs passivated by        hydrogen are structurally stable. 

\begin{figure*}[ht!]
	\includegraphics[width=18cm]{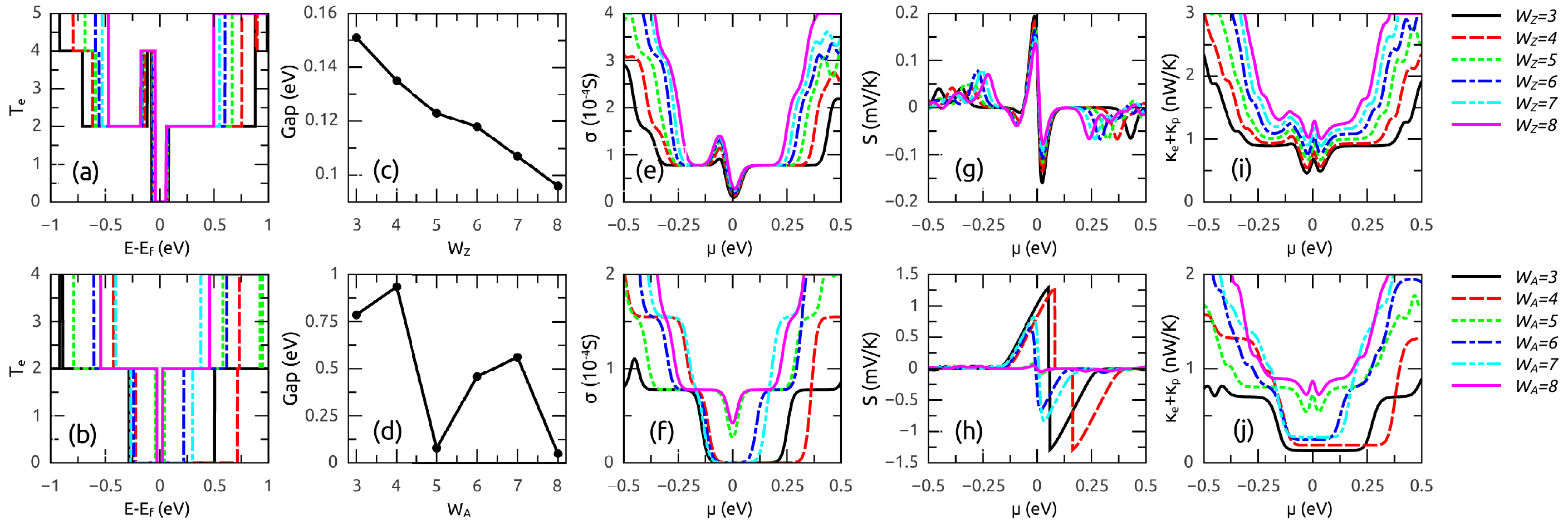} 
	\caption{(Color online) (a) and (b) Electron transmission coefficient as a function
of energy for Z-GeNRs and A-GeNRs with various ribbon width, respectively.
(c) and (d) Band gap of Z-GeNRs and A-GeNRs as a function of the ribbon width
$W_Z$ and $W_A$, respectively. (e) and (f) Electrical conductance, (g)
and (h) Seebeck coefficient, (i) and (j) electron and phonon thermal conductances for
Z-GeNRs and A-GeNRs versus chemical potential $\mu$, where the temperature is set 300K.} 
	\label{fig6} 
\end{figure*}

To calculate the figure of merit $ZT$, we begin with the electron transport
properties. Figures 6(a) and (b) show the transmission coefficient ${\cal T}_e$
as a function of the electron energy $E$ for both Z-GeNRs and A-GeNRs,
respectively. It can be seen that ${\cal T}_e$ exhibits a clear quantum
stepwise structure, due to opening and closing of elastic transmission
channels: notice that the jumps are quantized and equal to 2 due to the
electron spin. More interesting, a monotonously decreasing band gap is observed
in the Z-GeNRs with the increasing of the ribbon width (see Fig.
6(c)). This must be compared with the oscillatory behaviour we observe
for the A-GeNRs (see Fig. \ref{fig6} (d)). For the A-GeNRs, for the ribbon widths $W_A=3p$ and
$3p+1$ (where $p$ is positive integer), the gap is larger than that of the
ribbon width $W_A=3p+2$. By making use of the transmission probability, using
Eqs.~(\ref{conductance})-(\ref{seebeck}), we can calculate the electrical
conductance $\sigma$, Seebeck coefficient $S$ and electron contributed thermal
conductance $\kappa_e$. In Figs. \ref{fig6} (e) and (f), the electrical conductance as a
function of chemical potential is plotted for both Z-GeNRs and A-GeNRs,
respectively. It can be seen that the electrical conductance for zigzag
nano-ribbons gradually increases with the ribbon width, and there is a peak
corresponding the transmission step at $E\approx0.5$ eV. Around the
Fermi energy, the conductance vanishes due to the finite gap.
For the A-GeNRs, we find that the electrical conductance for the
ribbon with width $3p$ or $3p+1$ vanishes, while for the ribbon with width
$3p+2$, a non-zero dip is found. Interestingly, the conductance for all the
curves of A-GeNRs exhibits quantized plateau-like characteristics. 

In Fig.~\ref{fig6} (g) and (h) we report the Seebeck coefficient as a function
of the chemical potential $\mu$. It can be seen from Fig.~\ref{fig6}(g) that
$S$ presents two peaks around the position of the chemical potential needed to
overcome the gap. Moreover the two peaks show different sign with positive and
negative values. This behavior indicates the different carrier transport: the
positive sign in the region of $\mu<0$ corresponds to hole transport, while the
negative at $\mu>0$ corresponds to electron transport. In addition the absolute
value of the peak of the Seebeck coefficient decreases with increasing $W_Z$.
In the case of A-GeNRs, it is found (see Fig. \ref{fig6} (h)) that for the
nano-ribbons with width $3p$ and $3p+1$, the two Seebeck coefficient peaks with
opposite sign can also be found centred around zero value of the chemical
potential. We note that for the nano-ribbons with width $3p+2$, the Seebeck
coefficient is very small due to the small electronic gap. In Figs.~\ref{fig6}
(i) and (j) the total thermal conductance for Z-GeNRs and A-GeNRs is depicted,
respectively. It can be seen that the thermal conductance for Z-GeNRs increases
with increasing the width of the nano-ribbon. By checking the variation of the
electrical and thermal conductances, $\sigma$ and $\kappa$, it is found that
corresponding to the dip position of the electrical conductance, the electric
thermal conductance (and therefore the total thermal conductance) shows a peak
which becomes sharper with increasing $W_A$. Moreover, a similar effect can
also be found in the A-GeNRs with width $3p+2$ as shown in Fig.~\ref{fig6}(j).

\begin{figure}[ht!]
	\includegraphics[width=8.6cm]{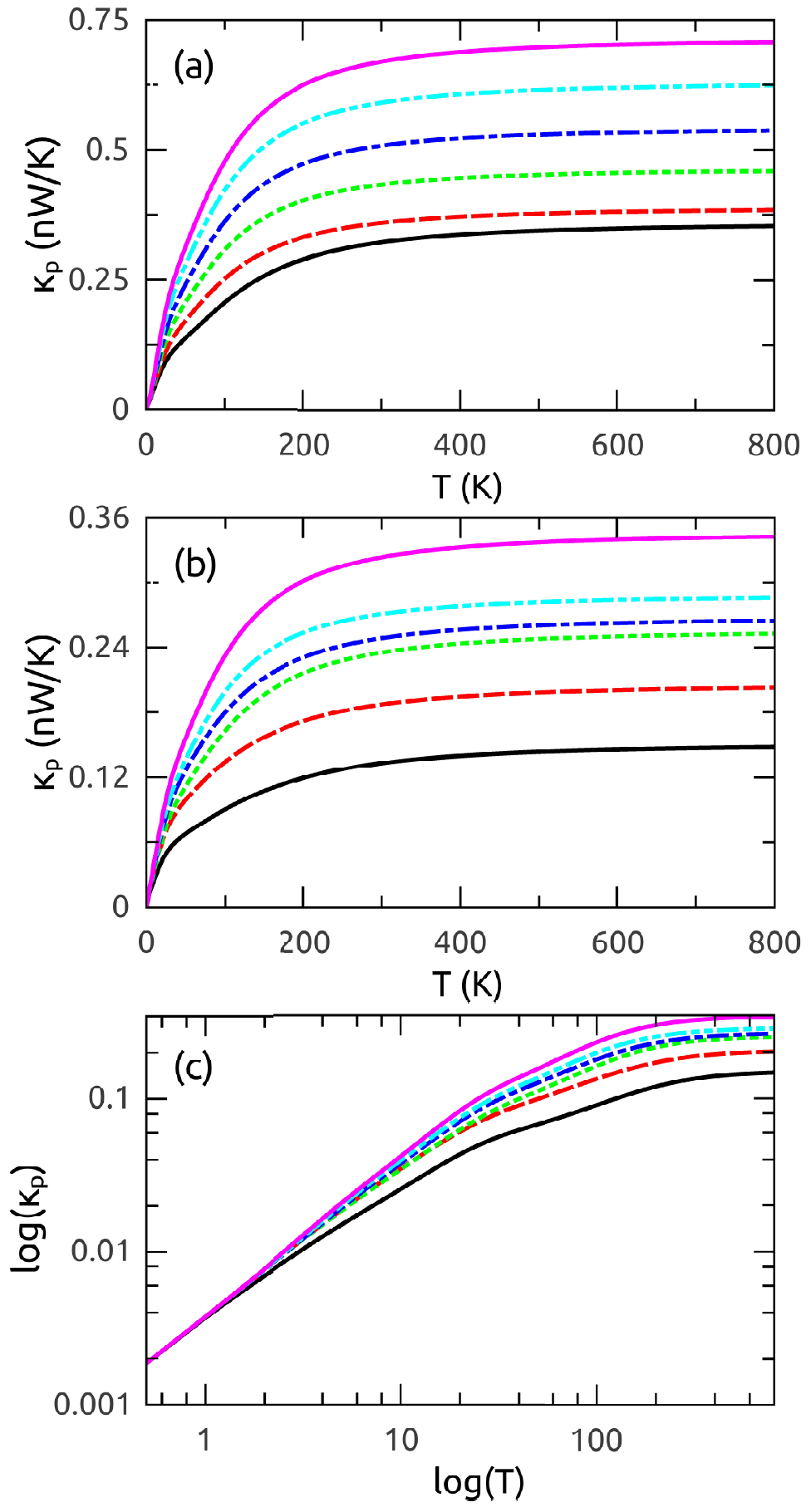} 
	\caption{(Color online) Phonon thermal conductance $\kappa_p$ of (a) Z-GeNRs
and (b) A-GeNRs with different ribbon width as a function of temperature. (c)
The logarithm of $\kappa_p$ for A-GeNRs as a function of logarithm of $T$, where
the linear behavior is shown as we expect according to Eq.~(\ref{quantumthermalc}).
Each curve corresponding to a specified ribbon width has the same meaning as in Fig. 6.}
	\label{fig7}
\end{figure}

To study the lattice thermal transport properties, the supercell approach is
utilised to calculate the phonon force constant and then the dispersion
relation is obtained by diagonalizing the corresponding dynamical matrix.\cite{Alfe2009} In
Figs. \ref{fig7} (a) and (b), the phonon thermal conductance $\kappa_p$ as a function of
temperature $T$ for both Z-GeNRs and A-GeNRs is plotted, respectively. It can be seen that
the phonon thermal conductance increases with increasing the temperature, and
finally reaches a constant value corresponding to the classical limit when
$T>400$ K. Moreover the thermal conductance for wide nano-ribbons exhibits a
higher value than that of the narrow nano-ribbons. This can simply be explained
by “counting” the number of phonon channels, because the wide nano-ribbons should
have more phonon channels contributing to the thermal transport. To show the
behaviour at low temperatures of the phonon thermal conductance, in Fig. \ref{fig7} (c),
we plot the logarithm of $\kappa_p$ versus the logarithm
of $T$. It can be seen that $\kappa_p$ shows a linear dependence on the
temperature at low $T$, $T<20$ K. At low temperature, for the one-dimensional
systems, the lattice thermal conductance is dominated by the low-frequency
acoustic phonons, and the Eq.~(\ref{thermalc}) can be recast as
\begin{equation} 
	\kappa_p(T)=\frac{4 k_B^2 T}{h}\int^{\infty}_{0} d\xi\, \xi^{2}\frac{e^{\xi}}
	{(e^{\xi}-1)^{2}}=\frac{2\pi k_B^2 T}{3\hbar},
    \label{quantumthermalc}
\end{equation}
where $\xi=\frac{\hbar\omega}{k_B T}$ and we have approximated the transmission
probability ${\cal T}_p(\omega)=4$ because of the sum rule. 
According to this approximation, it can be seen that
the phonon thermal conductance exhibits a linear dependence on $T$
in quasi-one-dimensional systems.\cite{Yamamoto2006} 

\begin{figure}[ht!] 
	\includegraphics[width=8.6cm]{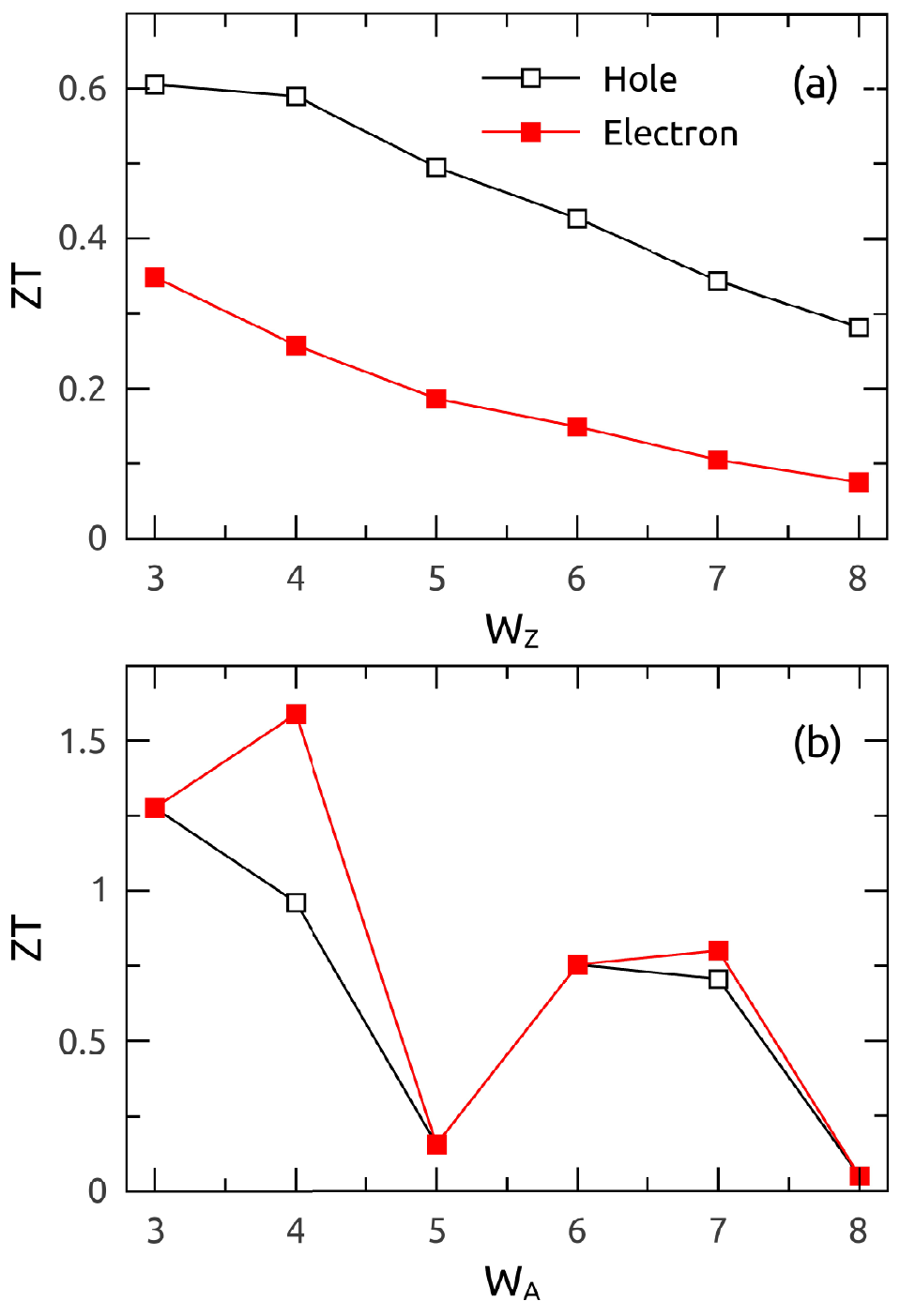}
    \caption{(Color online) Figure of merit $ZT$ at room temperature for (a) Z-GeNRs
and (b) A-GeNRs as a function of ribbon width $W_Z$ and $W_A$, respectively.
In black (square hollow points) we report the peak value of $ZT$ at negative
values of the chemical potential $\mu$ associated with the hole
transport, and in red (square full points) the peak value of $ZT$ associated with the
electron transport (positive $\mu$).}
	\label{fig8}
\end{figure}

By combining the results of the electron and phonon calculations, we can
finally investigate the thermoelectric efficiency of the GeNRs.
Figures 8(a) and (b) report the thermoelectric figure of merit $ZT$ as a
function of the ribbon width for both Z-GeNRs and A-GeNRs, respectively. Here
$ZT$ is the maximum value of the figure of merit with respect to the chemical
potential near the Fermi energy. It can be seen from Fig. \ref{fig8} (a) that at narrow
Z-GeNRs, $ZT$ for electron and hole is about 0.35 and 0.61, and then it
decreases with increasing the ribbon width. This effect can be explained by
the lessening of the Seebeck coefficient and growing of the thermal conductance
outweighing the increasing electrical conductance. Moreover, we observe from Fig.
8(b) for the A-GeNRs, that the $ZT$ for both electron and hole transport
coefficients show an oscillatory behavior. In the case of the nano-ribbons with
width $W_A=3p$ and $3p+1$, $ZT$ is larger than 1 for narrow nano-ribbons. In
particular for the ribbon width $W_A=4$, the $ZT$ reaches up to 1.63,
indicating a high thermoelectric conversion efficiency in these nano-structures. 

Our results are consistent with what has been found for SiNRs.\cite{Pan2012}
However, we would like to point out that from our
calculations the phonon thermal conductance of the small GeNR is never
negligible with respect to the electron thermal conductance, as instead has
been argued for the SiNRs in Ref.~\onlinecite{Pan2012}. We
believe this is an artefact of the classical methods used in
Ref.~\onlinecite{Pan2012}. Indeed, we expect that
the classical methods would not recover the linear behavior with temperature
of the phonon thermal conductance at small temperatures of these
quasi-one-dimensional systems in the ballistic thermal transport regime
as we do in our quantum simulations.
Moreover, the classical calculations should be valid for temperatures larger
than the Debye temperature, that for these systems can be estimated to be about
600 K for silicene nano-ribbons. At the same time, the quantum technique does
not include any inelastic effect and it is strongly limited in size, i.e., we
cannot consider a large supercell as instead is possible with classical
methods.\cite{Pan2012}

\subsection{Silicene nano-ribbons} 

\begin{figure}[ht!]
	\includegraphics[width=8.6cm]{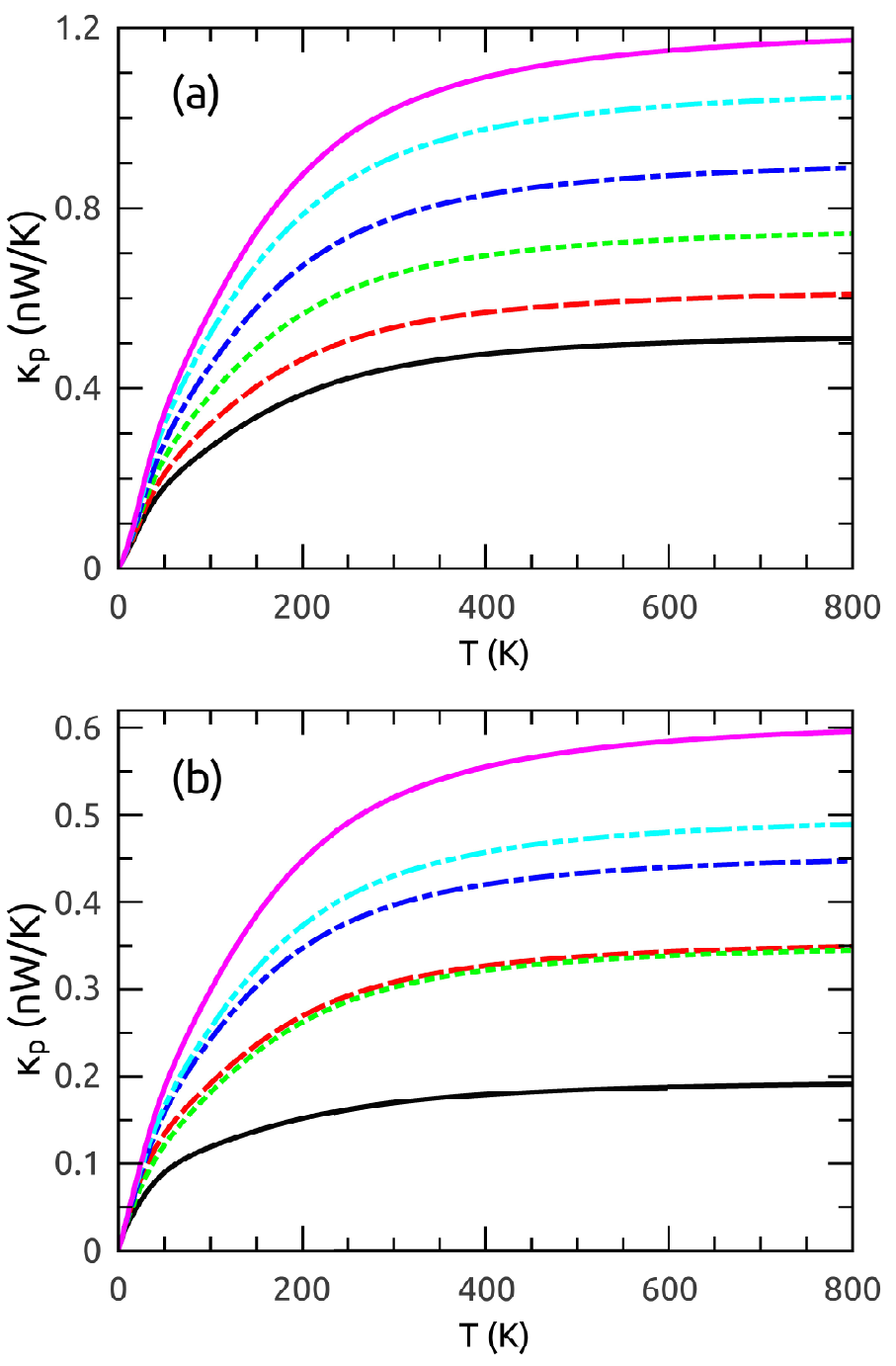}
    \caption{(Color online) Phonon thermal conductance $\kappa_p$ of (a) Z-SiNRs and
(b) A-SiNRs as a function of temperature, respectively, where each curve
corresponds to a specified ribbon width as shown in Fig. 6.}
	\label{fig9} 
\end{figure}

For completeness, and to have a direct comparison with the results available in
the literature,\cite{Pan2012} we have calculated, the figure of merit of SiNRs,
similar to the GeNRs we have investigated in the previous section. Here we
report only the phonon thermal conductance and the figure of merit. The
electron transport coefficients, $\sigma$, $S$, and $\kappa_e$ have shapes
similar to those in Fig.~\ref{fig6} and we do not show them again. We plot in
Figs. \ref{fig9} (a) and (b), the phonon thermal conductance $\kappa_p$ for
both zigzag- and armchair-edged SiNRs (Z-SiNRs and A-SiNRs) as a function of
temperature $T$, respectively. It is seen that the phonon thermal conductance
increases with the temperature, and finally reaches a steady value. For the
zigzag nano-ribbons, the thermal conductance increases gradually with the
ribbon width due to the increase in available phonon transport channels. As to
the armchair nano-ribbons, the thermal conductance is also increased except for
the ribbon width $W_A=3,~4$ whose values are tightly close (see Fig. \ref{fig9}
(b)).

\begin{figure}[ht!] 
	\includegraphics[width=8.6cm]{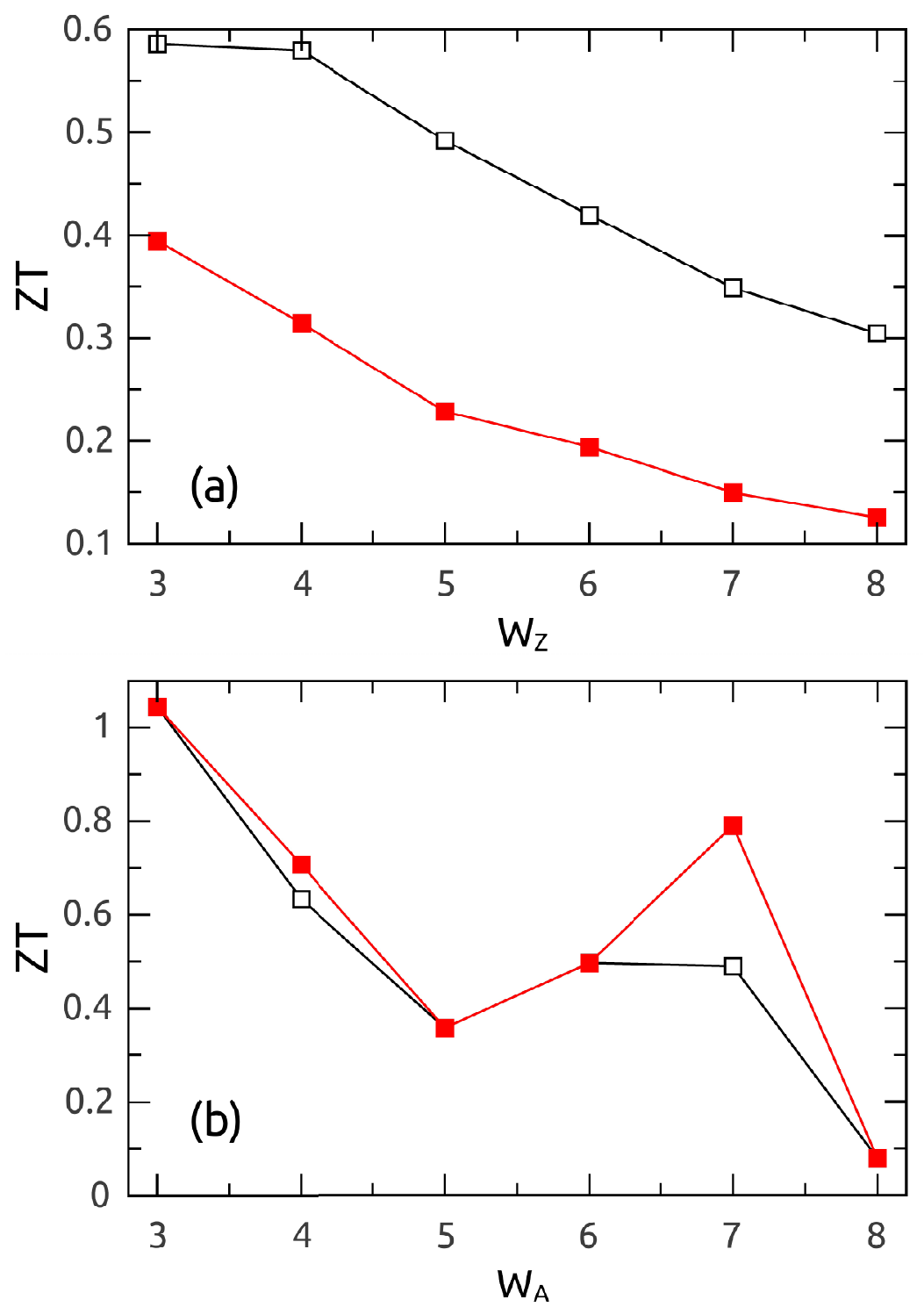}
    \caption{(Color online) Figure of merit $ZT$ at room temperature for (a) Z-SiNRs and
(b) A-SiNRs as a function of ribbon width $W_Z$ and $W_A$, where the black square hollow
points and the red square full points correspond to the hole and electron transport, respectively.}
	\label{fig10} 
\end{figure}

In Fig. \ref{fig10} the figure of merit for SiNRs as a function of ribbon width is shown. It can be
seen that the figure of merit for Z-SiNRs decreases with the increase of the ribbon width.
Moreover the $ZT$ for the hole transport is larger than that contributed from the electron transport.
The reason is due to the increased phonon thermal conductance and decreased electronic band gap.
For the armchair nano-ribbons, it is found from Fig. \ref{fig10} (b) that
the figure of merit at narrow ribbon is quite large, about 1.04.
With the increase of the ribbon width, the $ZT$ decreases overall and exhibits
an oscillatory behavior. 

\section{Silicon-germanium hetero-structures}
\label{1DSiGe}

We have shown that the Si and Ge nano-ribbons can have a substantial figure of
merit, which is slightly above 1. On the other hand, we would like to explore
the possibility of improving on this result by nano-structuring these
nano-ribbons. Since Si and Ge nano-ribbons do share similar electronic
properties, our first attempt is to investigate a nano-ribbon created by
alternating stripes of Si and Ge in the direction of the growth of the
nano-ribbon. Hopefully, their different masses would create a trap for the
phonon modes thus reducing the thermal conductance of the device and improving
the overall figure of merit $ZT$. We will show in the following sub-section
that this idea is working partially and we do have a modest increasing of $ZT$.
This is a limitation of our quantum method of calculating the thermal
conductance, since we are limited in the size of the supercell we can consider
for our calculations. Indeed, the low energy phonons responsible primarily for
the thermal transport have a wavelength that spans many supercells thus making
the chemical modulation ineffective as a phonon trap. To improve on this
result, we have therefore investigated the case where we randomly substituted
some Si atoms with Ge in the nano-ribbon crystal. After fully relaxing the
structure, we have however observed that also this nano-ribbon with randomly
distributed Si and Ge atoms does not work too much as a phonon trap, for
essentially the same reason of the perfect modulation: the Si and Ge randomly
distributed supercell is not large enough to confine the low energy phonon
modes. We checked this observation by using a tight-binding approximation to
calculate the phonon spectrum. This allows us to reach larger a supercell and
thus show that the phonon thermal conductance decreases due to the phonon
confinement in these random structures.

\subsection{Thermoelectric properties of the silicene-germanene nano-ribbons} 
\begin{figure}[ht!]
	\includegraphics[width=8.6cm]{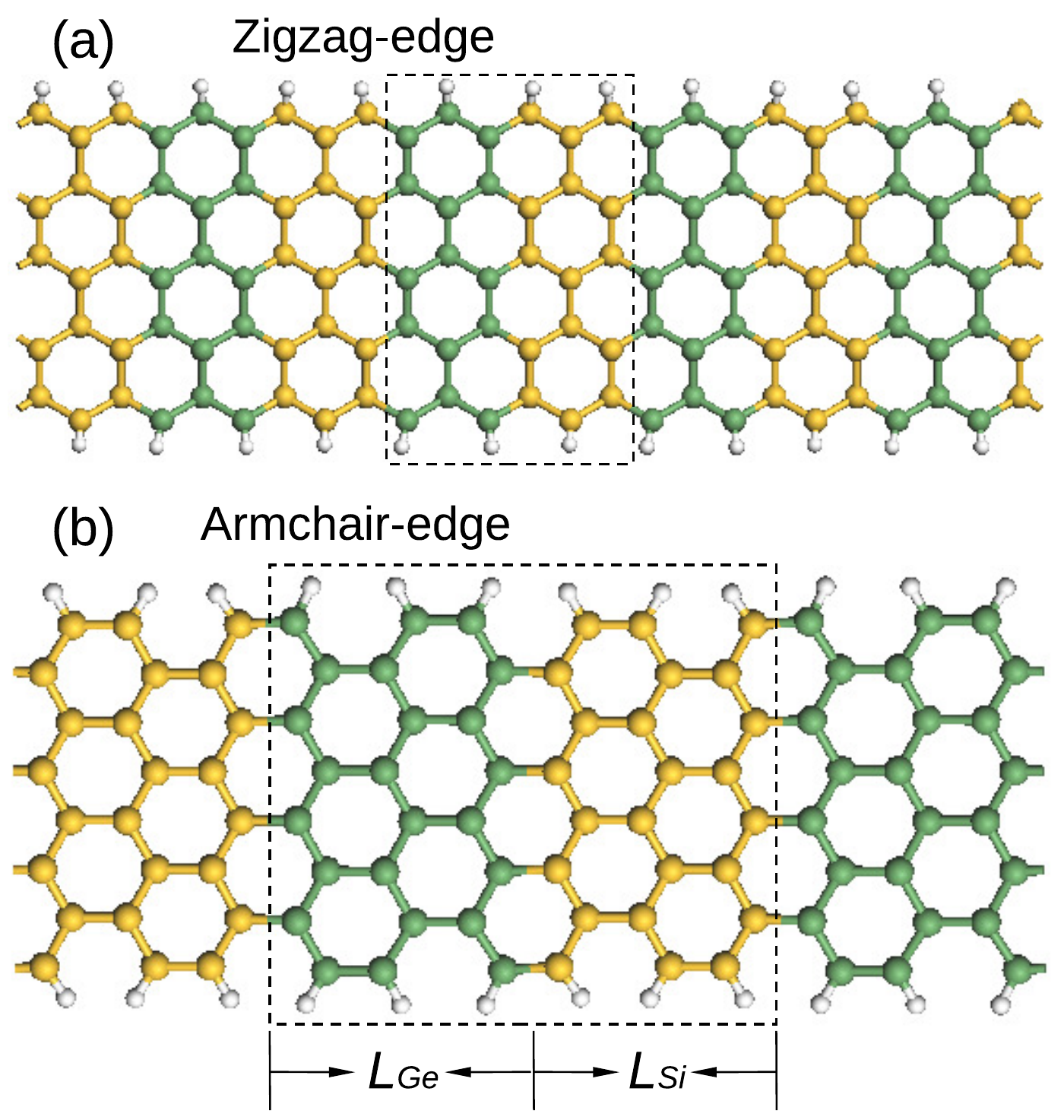} 
	\caption{(Color online) Geometrical structures of
(a) Z-SiGeNRs and (b) A-SiGeNRs, where the
line encloses a supercell along the ribbon axis and $L_{Si}$ and $L_{Ge}$
are the lengths of silicene and germanene stripes in the supercell, respectively.
Here we have chosen $L_{Si}=L_{Ge}=3$ and used the hydrogen to
passivate the ribbon edges.}
    \label{fig11} 
\end{figure}

In this section, we investigate the thermoelectric properties of
orderly-distributed hetero-structured silicene-germanene nano-ribbons
(SiGeNRs). After forming the structure, we have relaxed the atomic positions,
without taking into account any substrate. Figure \ref{fig11} shows the
optimized geometrical structures of zigzag- and armchair-edged SiGeNRs
(Z-SiGeNRs and A-SiGeNRs) passivated by hydrogen atoms, where the line encloses
a supercell along the ribbon axis. $L_{Si}$ and $L_{Ge}$ are the length of
silicene and germanene stripes in the supercell, respectively.

\begin{figure*}[ht!]
	\includegraphics[width=18cm]{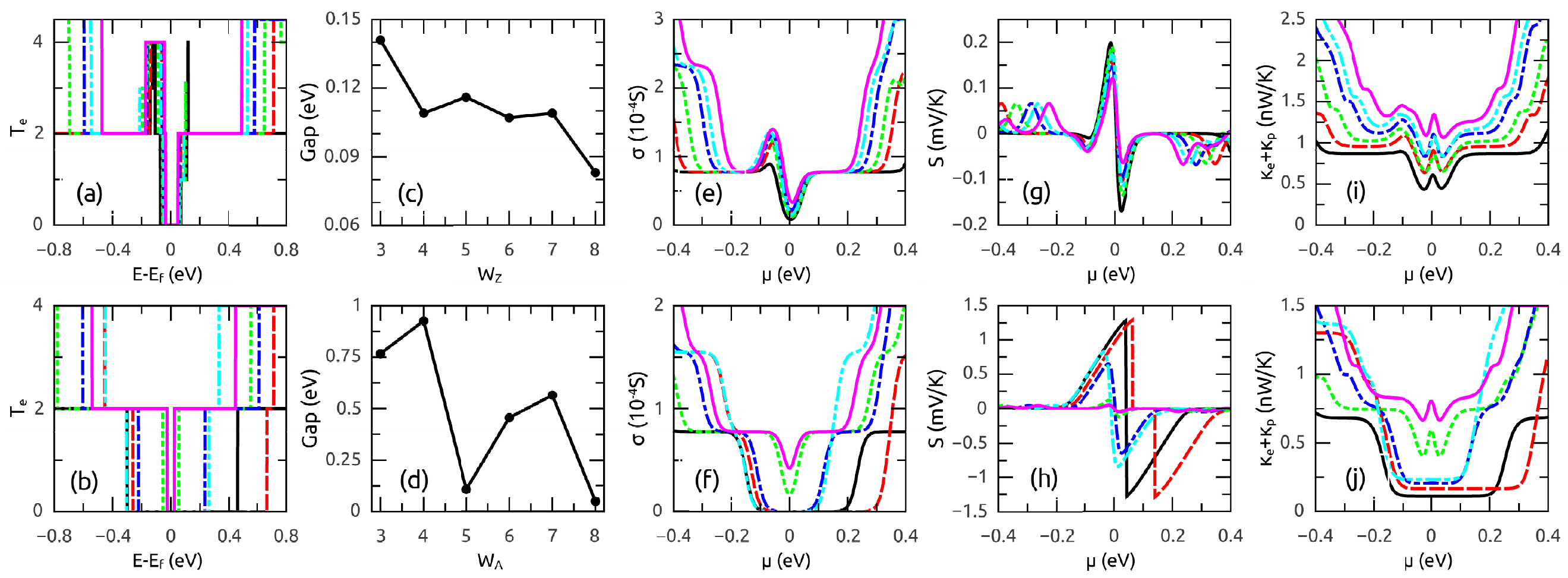} 
	\caption{(Color online) (a) and (b) Electron transmission coefficient as a function of
energy for Z-SiGeNRs and A-SiGeNRs, respectively. (c) and (d)
Band gap as a function of ribbon width $W_Z$ and $W_A$, respectively.
(e) and (f) Electrical conductance, (g) and (h) Seebeck coefficient, (i) and
(j) electron and phonon thermal conductances as a function of chemical
potential $\mu$ for Z-SiGeNRs and A-SiGeNRs, respectively, where we have set the
temperature T=300 K. Each curve in (a,b,e-j)
corresponding to a specified ribbon width endows the same meaning as in Fig. 6.}
	\label{fig12}
\end{figure*}

We begin with the case $L_{Si}=L_{Ge}=1$. In Figs. \ref{fig12} (a) and (b) we
report the transmission coefficient as a function of electron energy for
different width of the Z-SiGeNRs and A-SiGeNRs, respectively. It can be seen that
the transmission probability exhibits characteristic quantized steps and a band
gap is shown around the Fermi energy. Increasing the ribbon width, the band gap
for Z-SiGeNRs shows an oscillatory behaviour of decreasing amplitude from $W_Z=4$ to 7 (see Fig. \ref{fig12} (c)), while the gap for
A-SiGeNRs shows a strongly oscillatory behaviour as shown in Fig. \ref{fig12}
(d). When the ribbon width $W_A$ satisfies $3p$ either $3p+1$, a larger gap
appears than that of the nano-ribbons with width $3p+2$. This width dependence
of the band gap is similar to that of the A-GeNRs and A-SiNRs as we have
discussed in section \ref{1D}. Starting from this transmission function we can
now easily evaluate the Eq.~\ref{conductance}-\ref{thermalc} to obtain the
transport coefficients. In Figs. \ref{fig12} (e) and (f) we plot the electrical
conductance as a function of the chemical potential $\mu$ in the linear response. It
is found that the electric conductance for Z-SiGeNRs exhibits a peak and a dip
around $\mu=0$. As for the A-SiGeNRs, we show that the electrical conductance
is zero for the nano-ribbon with width $W_A=3p$, $3p+1$ due to the presence of
the larger band gap, while the conductance for the ribbon with width $W_A=3p+2$ has
a dip at $\mu=0$ where the conductance assumes a finite value. In Figs. 12 (g)
and (h), the Seebeck coefficient versus chemical potential is depicted. It is
found that in the Seebeck coefficient, around $\mu=0$ two peaks appear for both
Z-SiGeNRs and A-SiGeNRs with width $W_A=3p$, $3p+1$. The absolute value of the
peak for A-SiGeNRs is 1.4 mV/K, which is quite larger than the value of the
Z-SiGeNRs, indicating a quite high thermoelectric effect in this armchair-edged
nano-ribbons. On the other hand, for the armchair nano-ribbons with width
$3p+2$, the Seebeck coefficient is very small due to the very small gap present in these
systems. In Figs. \ref{fig12} (i) and (j) the total thermal conductance
$\kappa_e+\kappa_p$ including electron and phonon contributions is plotted. It
can be seen that $\kappa=\kappa_e+\kappa_p$ for Z-SiGeNRs exhibits a peak, while for
the A-SiGeNRs with widths $3p$ and $3p+1$, it has a plateau in the energy region
around $\mu=0$, mostly due to the phonon thermal transport. As for the nano-ribbon
with width $3p+2$, the thermal conductance reaches a local maximum on account
of a local maximum of the electron heat contribution at $\mu=0$. 

\begin{figure}[ht!]
	\includegraphics[width=8.6cm]{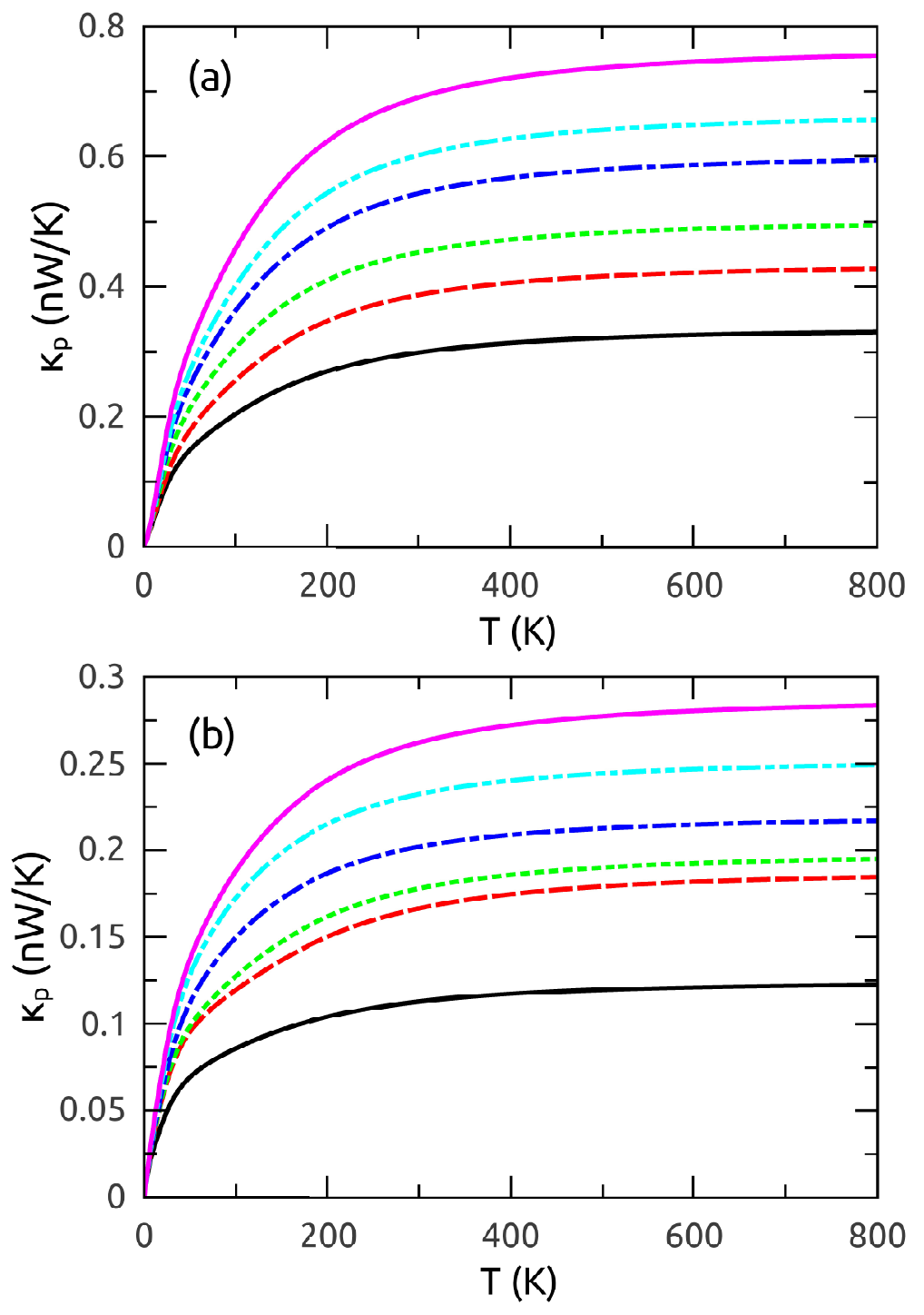} 
	\caption{(Color online) Phonon thermal conductance $\kappa_p$ of
(a) Z-SiGeNRs and (b) A-SiGeNRs as a function of temperature, respectively,
where each curve corresponding to a specified ribbon width shares the same
meaning as shown in Fig. 6.}
	\label{fig13}
\end{figure}

Figures \ref{fig13} (a) and (b) show the phonon thermal conductance $\kappa_p$ for both
Z-SiGeNRs and A-SiGeNRs as a function of temperature for different nano-ribbon
widths, respectively. It can be seen that the phonon thermal conductance
$\kappa_p$ increases gradually with increasing the temperature and finally
reaches a plateau at $T>400$ K. This value is compatible with the value reported in the classical
thermal transport theory. By comparing
Fig.~\ref{fig13} with Figs.~\ref{fig7} and \ref{fig9}, we find that the thermal
conductance of SiGeNRs is between the value of GeNRs and SiNRs, i.e., the thermal conductance
of SiGeNRs is larger than that of GeNRs, while smaller than that of SiNRs.
Similar to the case of GeNRs or SiNRs,
at low temperature region, the linear dependence of the thermal
conductance on the temperature is still observed, in agreement with Eq.~(\ref{quantumthermalc}). 

\begin{figure}[ht!]
	\includegraphics[width=8.6cm]{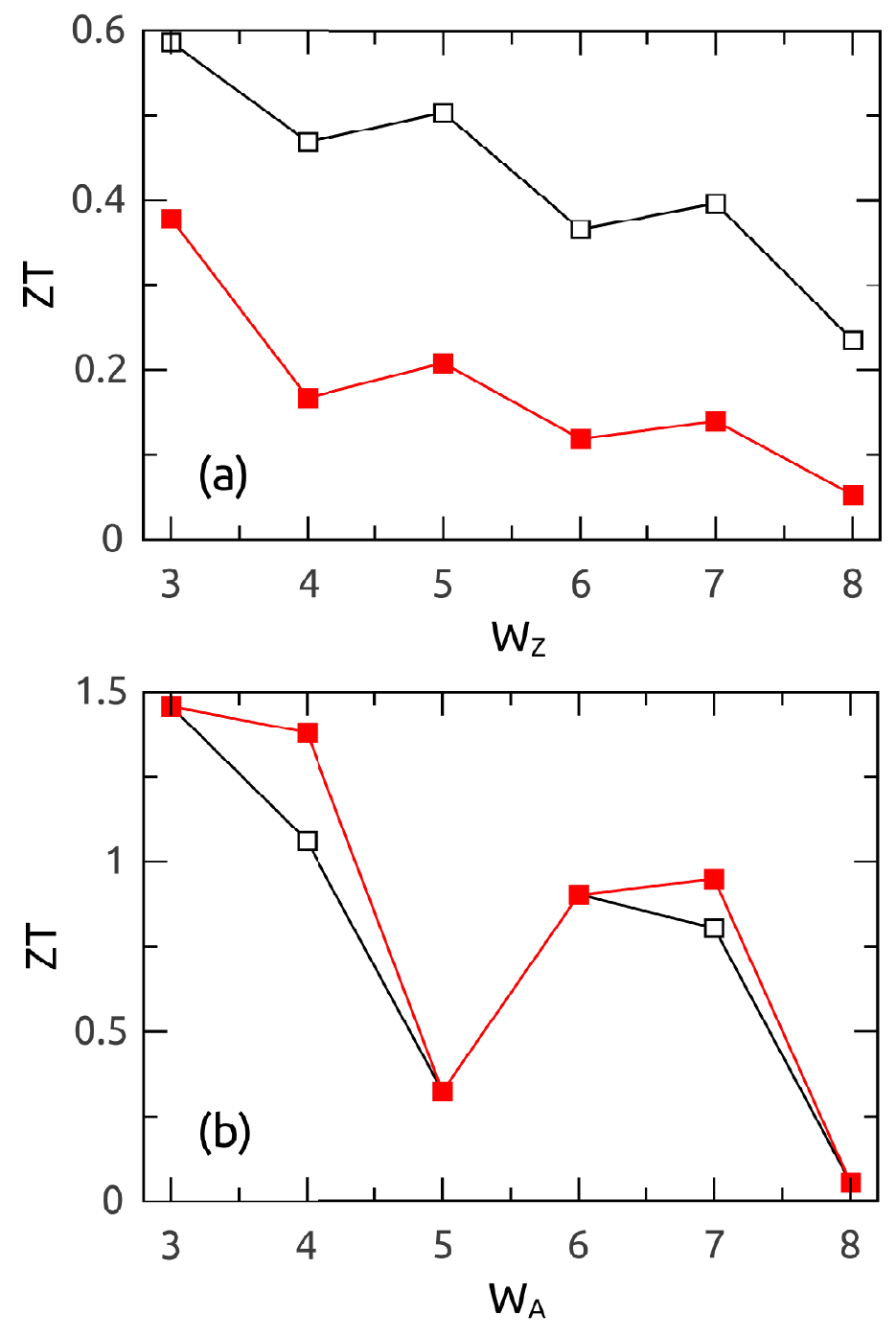}
	\caption{(Color online) Figure of merit $ZT$ at room temperature for (a) Z-SiGeNRs and
(b) A-SiGeNRs as a function of nano-ribbon with width $W_Z$ and $W_A$,
where the hollow and full points correspond to the hole and electron transport, respectively.}
    \label{fig14} 
\end{figure}

In Figs.~\ref{fig14} (a) and (b), we report the figure of merit $ZT$ of
both Z-SiGeNRs and A-SiGeNRs as a function of
ribbon widths $W_Z$ and $W_A$, respectively. It is found that maximum value of the figure of
merit for Z-SiGeNRs appears in the narrowest nano-ribbon, which is about 0.59 corresponding
to the hole transport. While for the electron transport, the corresponding $ZT$ is about 0.38.
As to the armchair-edged nano-ribbon with width $W_A=3$,
the $ZT$ is found to be 1.46 for both the hole and electron transport
(see Fig.~\ref{fig14} (b)). With the increase of the
ribbon width, the figure of merit shows an oscillatory behavior
reminiscent of the different properties of the nano-ribbons with
different widths. The amplitude of the “oscillation” however decreases quite
rapidly with increasing the ribbon width. This is mostly due to the rapid
increasing of the phonon thermal conductance with $W_A$. In particular, the $ZT$ is
very small in the case of nano-ribbon with width $3p+2$ due to the small
Seebeck coefficient as shown in Fig.~\ref{fig12} (h). 

\subsection{Component modulation of the thermoelectrics in the silicene-germanene nanoribbons} 

\begin{figure*}[ht!] 
	\includegraphics[width=18cm]{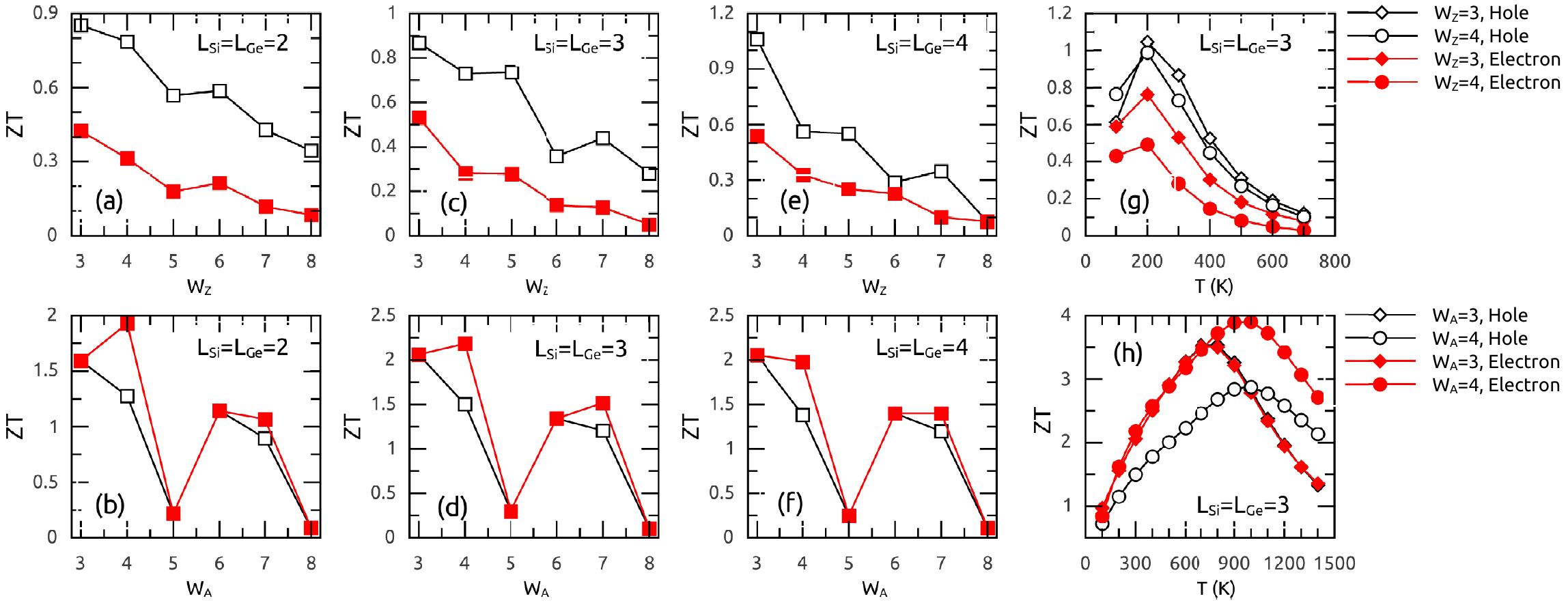} 
	\caption{(Color online) Figure of merit $ZT$ at T=300K for Z-SiGeNRs and
A-SiGeNRs as a function of ribbon width $W_Z$ and $W_A$ under different
component lengths of silicene and germanene stripes: (a) and (b)
$L_{Si}=L_{Ge}=2$, (c) and (d) $L_{Si}=L_{Ge}=3$, (e) and (f)
$L_{Si}=L_{Ge}=4$, respectively. (g) and (h) Figure of merit as a function of
temperature for Z-SiGeNRs and A-SiGeNRs with the corresponding ribbon width 3
and 4, where the lengths of silicene and germanene stripes in the supercell are
$L_{Si}=L_{Ge}=3$. The hollow and full points correspond to the hole and
electron transport, respectively
}
	\label{fig15} 
\end{figure*}

In Fig.~\ref{fig15}, we investigate the thermoelectric properties of SiGeNRs by
modulating the component lengths of silicene and germanene stripes in the
supercell. In the following, the total length of the supercell is given by
$L_S=L_{Ge}+L_{Si}$.
Figures~\ref{fig15} (a) and (b), (c) and (d), (e) and (f) show the figure of merit $ZT$
at room temperature for Z-SiGeNRs and A-SiGeNRs as a function of ribbon width
for $L_{Si}=L_{Ge}=2$, $3$ and $4$, respectively. It is found that the maximum $ZT$ for hole
and electron transport in the case of the Z-SiGeNRs is 0.85 and 0.42 for $L_{Si}=L_{Ge}=2$,
0.87 and 0.53 for $L_{Si}=L_{Ge}=3$, and 1.06 and 0.54 for $L_{Si}=L_{Ge}=4$, respectively.
For armchair nano-ribbons, the maximum of
$ZT$ for $L_{Si}=L_{Ge}=2$ is about 1.93, while the maximum $ZT$ for
$L_{Si}=L_{Ge}=3$ or $4$ is about 2.18 and 2.06, respectively. With the increase
of the ribbon width, the overall figure of merit decreases
for both Z-SiGeNRs and A-SiGeNRs with width $W_A=3p$ and
$3p+1$. As to the nano-ribbon with width $3p+2$, the figure of merit is quite
small compared to the ribbons with width $3p$ or $3p+1$. We found that the Seebeck coefficient for nano-ribbon
with width $3p+2$ is very small due to the small band gap, in agreement with our 
analysis of the system with $L_{Si}=L_{Ge}=1$.
Figures \ref{fig15} (g) and (h) show the figure of merit as a function of temperature for Z-SiGeNRs
with width $W_Z=3,~4$ and A-SiGeNRs with width $W_A=3,~4$, respectively. It can be seen that
the figure of merit increases and then decreases with increasing the temperature.
The maximum $ZT$ for Z-SiGeNRs is about 1.05 at $T\approx 200$ K, and the maximum $ZT$
for A-SiGeNRs is about 3.91 at $T\approx 1000$ K. 

We wish to point out that the $ZT$ of these systems is larger than that for
the pure A-GeNRs or A-SiNRs. This means that nano-structuring can improve the
overall energy conversion efficiency. On the other hand, the modest increase in
$ZT$ for these nano-ribbons shows how this nano-structuring is not effective
in blocking the phonon modes. We should reach larger $L_{Si}$ and
$L_{Ge}$, in order to achieve an efficient trapping of the low energy phonon modes, as we will discuss briefly at the end of next sub section. 

\subsection{Disorder effect on the thermoelectrics of silicene-germanene nano-ribbons}
\begin{figure}[ht!] 
	\includegraphics[width=8.6cm]{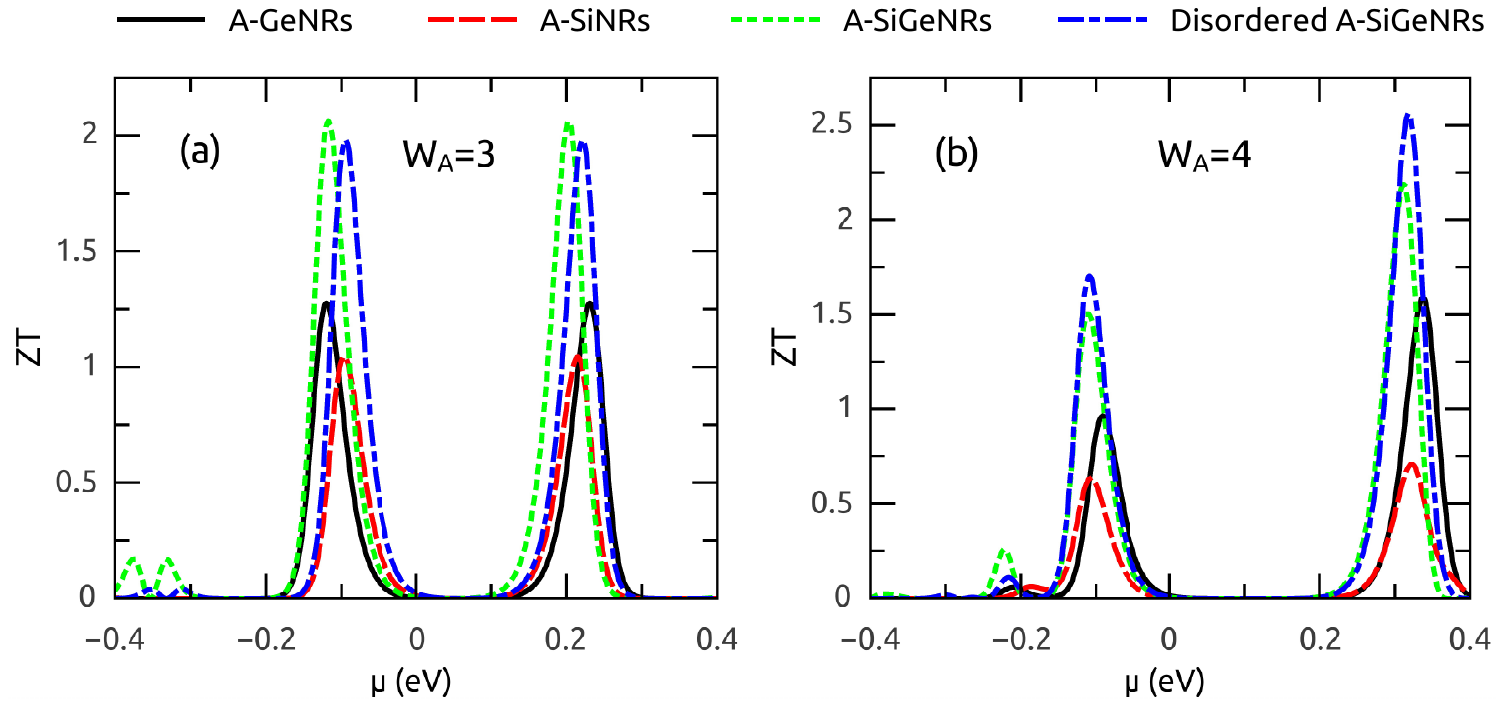}
	\caption{(Color online) Figure of merit $ZT$ at T=300 K for A-GeNRs, A-SiNRs, A-SiGeNRs
and disordered A-SiGeNRs with ribbon width (a) $W_A=3$ and (b) $W_A=4$ as a
function of chemical potential $\mu$, where we have taken the supercell length
$L_S=L_{Ge}+L_{Si}=6$, respectively.}
	\label{fig16}
\end{figure}

In the above discussions, the Si and Ge atoms in the nano-ribbons are orderly
distributed along the growth direction. Here we consider the case in which Si
and Ge atoms randomly occupy with equal probability the sites of the lattice in
Fig.~\ref{fig11}. The length of the supercell is $L_S=6$ and the number of Si
and Ge atoms in the supercell are taken the same. Since the armchair
nano-ribbons show the most promising values of the figure of merit, in Figs.
\ref{fig16} (a) and (b) we report the figure of merit $ZT$ as a function of the
chemical potential $\mu$ for disordered A-SiGeNRs with the ribbon width $W_A=3$
and $4$, respectively. As a comparison, we have also plotted the figure of
merit for A-GeNRs, A-SiGRs and A-SiGeNRs. It can be seen that the maximum
figure of merit for disordered A-SiGeNRs and ordered A-SiGeNRs is nearly twice
of the value of clean A-GeNRs and A-SiGRs. The maximum $ZT$ for disordered and
ordered A-SiGeNRs with width $W_A=3$ is about 2 for both electron and hole
transport corresponding to the positive and negative chemical potential, while
the maximum $ZT$ for the ribbon width $W_A=4$ is 2.18 and 2.56 for electron and
1.5 and 1.8 for hole transport, respectively. The principal reason of the
enhanced thermoelectric efficiency comes from the reduced phonon thermal
conductance, since the electronic properties are slightly affected by the
randomness of the atomic positions. Again, due to the small size of the
supercell we can consider with ab-initio techniques, phonon confinement is not
efficient, and therefore the thermal conductance of disordered and ordered
A-SiGeNRs is only slightly reduced with respect to the clean Si and Ge system
as shown in comparing Figs.~\ref{fig15} and~\ref{fig16}. For the same reason,
the thermal conductance of the random structure is similar to the one of the
silicene-germanene hetero-structures as expected.

\begin{figure}[ht!] 
	\includegraphics[width=8.6cm]{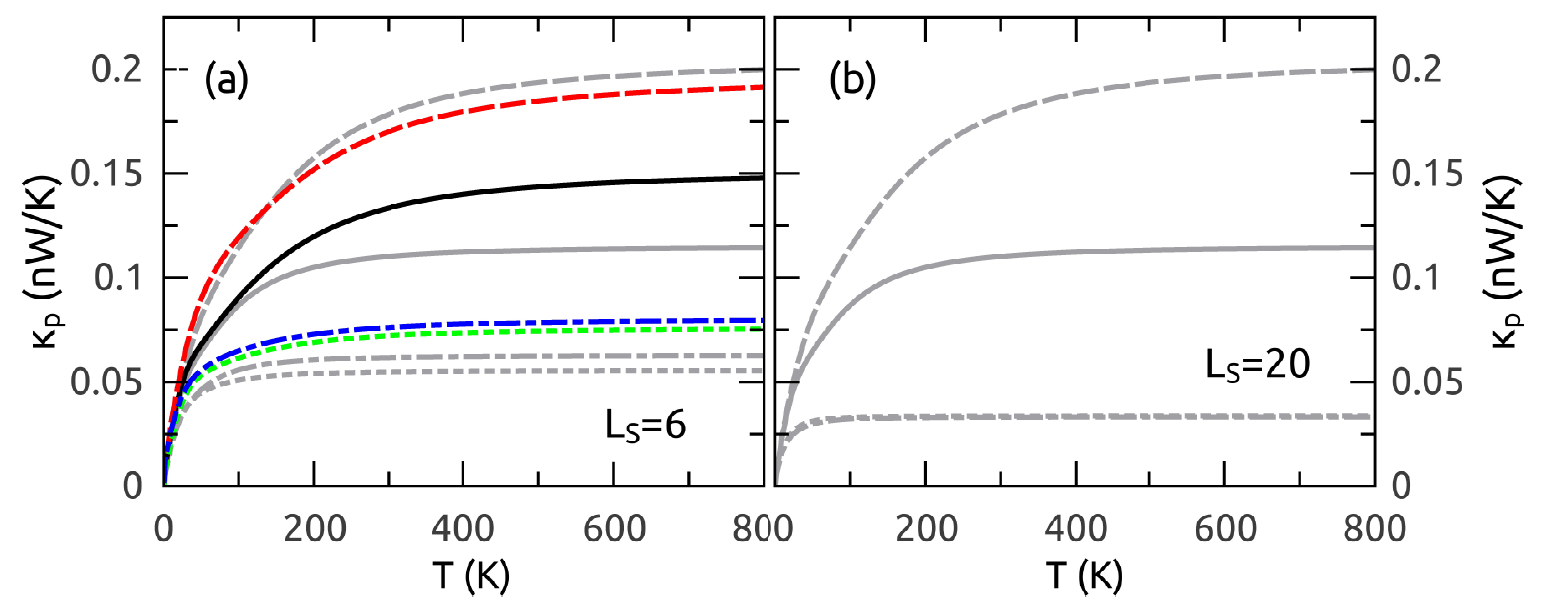}
	\caption{(Color online) Phonon thermal conductance of various armchair
nano-ribbons with the ribbon width $W_A=3$, where the supercell length $L_S$
for A-SiGeNRs and disordered A-SiGeNRs is (a) $L_S=6$ and (b) $L_S=20$,
respectively. The curves in color are calculated from ab-initio and the other
curves in gray are calculated from tight-binding. The different curves share
the same meaning as those in Fig.~\ref{fig16}.}
	\label{fig17}
\end{figure}

To present a proof that a large supercell can effectively further reduce the
phonon thermal conductance in the SiGe hetero-structures, we use a
semi-classical tight-binding method to investigate the lattice thermal
transport properties. To obtain the atomic force constant of the system, the
Keating potential is used,\cite{Rucker1995,Wang2008a,Davydov2012} which is
given by
\begin{eqnarray} 
	U&=&\frac{1}{2} k_r \sum_{i,j}\left(\mathbf{R}_{i,j}^2-\mathbf{r}_{i,j}^2\right)^2\nonumber\\
	&&+\frac{1}{2} k_\theta \sum_{i,j,k\neq j}
	\left(\mathbf{R}_{i,j}\cdot\mathbf{R}_{i,k}-\mathbf{r}_{i,j}\cdot\mathbf{r}_{i,k}\right)^2,
    \label{keating}
\end{eqnarray}  
where $\mathbf{R}_{i,j}$ and $\mathbf{R}_{i,k}$ are the equilibrium position
vectors connecting atom $i$ with $j$ and $k$, $\mathbf{r}_{i,j}$ and
$\mathbf{r}_{i,k}$ are the corresponding position vectors after deformation,
respectively. The bond stretching and bending force parameters $k_r$ and
$k_\theta$ for silicene in Eq.~(\ref{keating}) are 7.2186$\times10^{20}$ N/m$^3$
and 1.5225$\times10^{20}$ N/m$^3$. These constants can be obtained from the force constants of graphene.\cite{Davydov2010,Davydov2012} In our case we have used
\begin{eqnarray}
	k_r=\frac{2\alpha}{d},~k_\theta=\frac{\beta}{d}
\end{eqnarray}
where $\alpha=81$ N/m$^2$ and $\beta=34$ N/m$^2$ for
silicene\cite{Davydov2010,Davydov2012} and $d$ is the equilibrium distance of
the Si atoms in the silicene structure, that we have calculated as $d=2.244$
\AA. We have then fine tuned the values of $k_r$ and $k_\theta$ to improve the
agreement between the phonon spectrum (not shown) calculated via ab-initio and
the one calculated within the tight-binding approximation. For the parameters
of germanene, we roughly estimate $k_r=5.3469\times10^{20}$ N/m$^3$ and
$k_\theta=1.2516\times10^{20}$ N/m$^3$ through comparing the force constant
ratio of this 2D system with the bulk silicon and germanium
crystals.\cite{Rucker1995} We have again fine tuned these values to improve the
agreement between the ab-initio and tight-binding phonon spectra. For the force
parameters between Si and Ge atoms in the hybrid structures, we take their
average value. As to the Si-H and Ge-H interactions, we take ten percent of the
corresponding Si-Si and Ge-Ge interactions, accordingly. Based on this Keating
model and combined with the non-equilibrium Green's function technique, we can
calculate the phonon transmission probability and thus the thermal transport
properties (see details in Ref.~\onlinecite{Yang2012}). Figures \ref{fig17} (a)
and (b) show the phonon thermal conductance calculated from tight-binding (gray
lines) for A-GeNRs, A-SiNRs, A-SiGeNRs and disordered A-SiGeNRs, where the
ribbon width $W_A=3$. To check how reliable is the tight-binding calculations,
we report together the thermal conductance calculated from ab-initio (color
lines in Fig. \ref{fig17}(a)). It can be seen that the thermal conductance
obtained from tight-binding and ab-initio are quite close, especially in the
low temperature region. In addition, it is found that the phonon thermal
conductance in the case of A-SiGeNRs and disordered A-SiGeNRs is drastically
decreased compared to the pure A-SiNRs and A-GeNRs. With further increasing the
length of the supercell, the phonon thermal conductance is decreasing as shown
in Fig.~\ref{fig17}(b).

\begin{figure}[ht!] 
	\includegraphics[width=8.6cm]{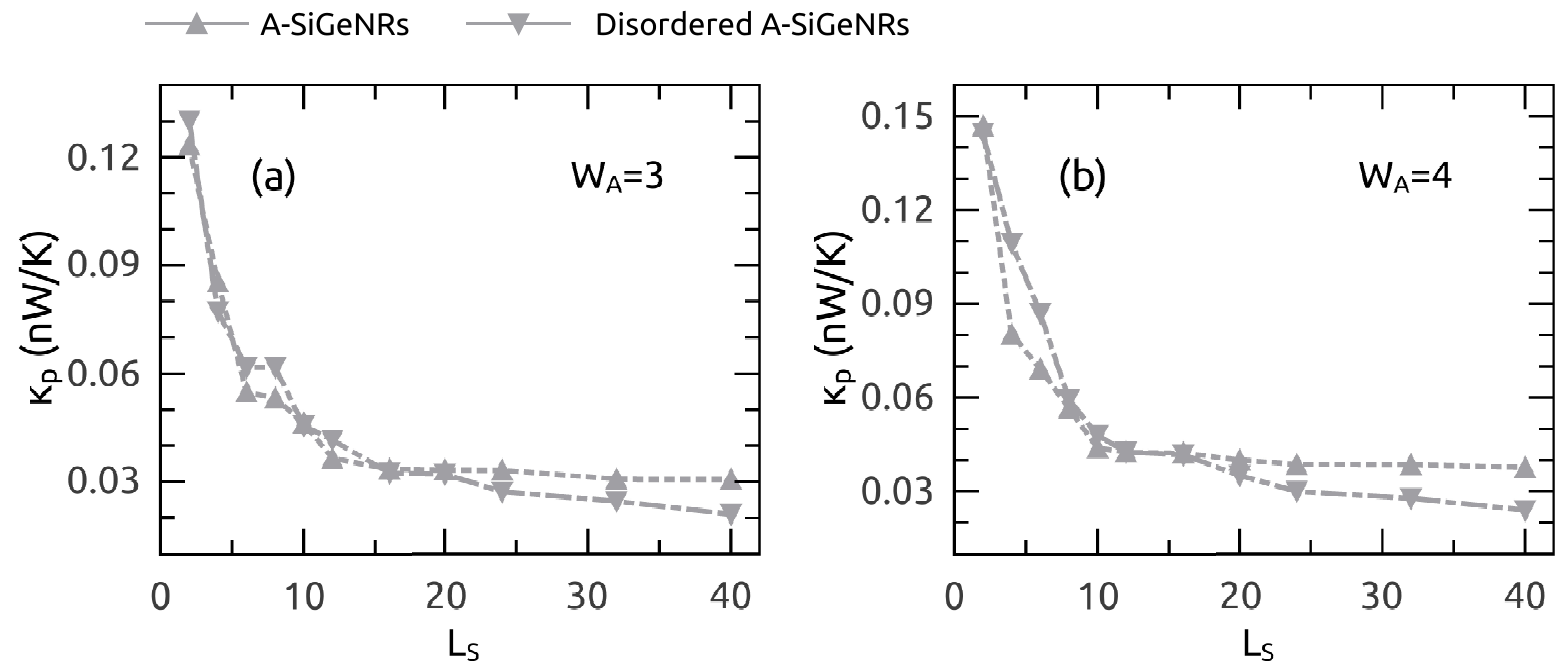}
		\caption{Phonon thermal conductance $\kappa_p$ at T=300 K calculated
from tight-binding as a function of supercell length $L_S$
corresponding to the ribbon width (a) $W_A=3$ and (b) $W_A=4$, respectively. }
	\label{fig18}
\end{figure}

In Fig.~\ref{fig18}, the phonon thermal conductance at T=300 K for both
A-SiGeNRs and disordered A-SiGeNRs as a function of supercell length $L_S$ is
investigated, where $L_S$ is defined as the sum of the length of silicene and
germanene stripes. It can be seen that the phonon thermal conductance decreases
with increasing the length of the supercell, and the $\kappa_p$ for both
A-SiGeNRs and disordered A-SiGeNRs are close to each other. This indicates that
the larger supercell in the silicene-germanene hetero-structures can
effectively constrain the phonon transport and that the disordered
heterostructure becomes more efficient in confining phonons only at large unit
cell lengths. Through checking the transmission probability (not shown), it is
found that the weight of the transmission probability is gradually moved to the
low frequency region thus decreasing the total energy flow.

\section{Conclusions}
\label{Conclusions} 
In summary, we have performed first-principle calculations of the
thermoelectric coefficients of both two-dimensional silicene and germanene as
well as for Si and Ge nano-ribbons. We have also considered hetero-structures
of Si and Ge stripes to form a nano-ribbon, in the attempt to quench the phonon
dynamics and thus increasing the figure of merit. These systems can be good
thermoelectric materials if they can be reliably produced. Also, being based on
Si and Ge, we expect these devices to be easily interfaced with the modern
electronic systems, a distinct advantage with respect to other materials which
have shown poor integrability with the actual technology.

The figure of merit for thermoelectric energy conversion of Si and Ge low
dimensional systems is quite high, in the range 1 to 2 at room temperature,
considering that we have investigated pristine systems where phonons are not
confined. For the silicene and germanene systems we have considered both
distorted silicene/germanene grown on a silver surface and free-standing
silicene/germanene. In these cases we have found the highest figure of merit is
about 0.81 for the distorted silicene. It is important to point out that Si/Ge
nano-sheets grown on a Ag surface show different electrical properties
according to the lattice matching: we have considered a 3$\times$3 Si lattice
on a 4$\times$4 Ag substrate since this induces no stress at the supercell
edges and when the Ag substrate is removed, the distorted silicene has a finite
band gap. It is indeed clear from our calculations that in order to increase
the figure of merit, we need to have a small gap semiconductor since this
maximizes the Seebeck coefficient.

Our attempts to quench the phonon dynamics have been hindered by the small
scale of the supercell we can calculate with our ab-initio techniques. We could
in principle go beyond these limitations by using other classical tools like,
e.g., molecular dynamics. However, especially for the nano-ribbons, in our
tests (not reported here) these tools have proven unable to recover the quantum
of thermal conductance at small temperatures. We have therefore chosen to test
the phonon confinement with semi-empirical techniques, e.g., a tight-binding
calculation of the phonon thermal transport for large supercell. We report that
the thermal conductance is effectively reduced by about 50\% in going from a
supercell made of 6 units to a supercell made of 20 units.

\acknowledgements
We thank Y. Pouillon and A. Iacomino for providing some computational help. We
acknowledge financial support from CONSOLIDER INGENIO 2010: NANOTherm
(CSD2010-00044), Diputacion Foral de Gipuzkoa (Q4818001B), the European
Research Council Advanced Grant DYNamo (ERC-2010-AdG-267374), Spanish Grants
(FIS2010-21282-C02-01), Grupos Consolidados UPV/EHU del Gobierno Vasco
(IT578-13), Ikerbasque, and MAT2012-33483. Computational time was granted by
i2basque and BSC “Red Espa\~nola de Supercomputacion”.

\bibliographystyle{apsrev4-1}
\bibliography{library2}

\end{document}